\documentclass[12pt]{article}
\usepackage{amstext,amsfonts,amsmath,amssymb,theorem}
\usepackage{graphicx}
\usepackage[left=2.5cm,top=3cm,right=2.5cm,foot=1.5cm]{geometry}
\allowdisplaybreaks
\usepackage{hyperref}
\hypersetup{%
  pdftitle   = {Contractions of Filippov algebras},
  pdfauthor  = {J. A. de Azc\'arraga, J. M. Izquierdo, M. Pic\'on}, }
\begin{document}
\begin{flushright}
FTUV-10-0901 $\;$ IFIC/10-29\\
September 1st, 2010 \\
(a few misprints corrected Dec. 13)\\
To appear in J. Math. Phys.
\end{flushright}
\vskip 2cm

\begin{center}
{\Large\bf Contractions of Filippov algebras}

\vskip 2cm
 {Jos\'e A. de Azc\'{a}rraga, \\
{\it Dept. of Theoretical Physics and IFIC (CSIC-UVEG), University of Valencia, \\
 46100-Burjassot (Valencia), Spain}\\
 \vskip .5cm
 Jos\'e M. Izquierdo,\\
{\it Dept. of Theoretical Physics, University of Valladolid, \\
47011-Valladolid, Spain}
 \vskip .5cm
  Mois\'es Pic\'{o}n \\
{\it Dept. of Theoretical Physics and IFIC (CSIC-UVEG), University of Valencia, \\
 46100-Burjassot (Valencia), Spain}\\
{\it C.N.Yang Institute for Theoretical Physics, Stony Brook University, \\
Stony Brook, NY 11794-3840,USA}}
\end{center}

\vskip 1.5cm

\begin{abstract}
We introduce in this paper the contractions $\mathfrak{G}_c$ of
$n$-Lie (or Filippov) algebras $\mathfrak{G}$ and show that they have a
semidirect structure as their $n=2$ Lie algebra counterparts.
As an example, we compute the non-trivial contractions of
the simple $A_{n+1}$ Filippov algebras. By
using the \.In\"on\"u-Wigner and the generalized Weimar-Woods
contractions of ordinary Lie algebras, we compare (in the
$\mathfrak{G}=A_{n+1}$ simple case) the Lie algebras
Lie$\,\mathfrak{G}_c$ (the Lie algebra of inner endomorphisms
of $\mathfrak{G}_c$) with certain contractions
$(\mathrm{Lie}\,\mathfrak{G})_{IW}$ and
$(\mathrm{Lie}\,\mathfrak{G})_{W-W}$ of the Lie algebra
Lie$\,\mathfrak{G}$ associated with $\mathfrak{G}$.
\end{abstract}.

\section{Introduction}

In 1985, Filippov \cite{Filippov, Filippov:98} initiated the study
of certain linear algebras (called $n$-Lie algebras by him) endowed
with a completely antisymmetric bracket with $n$ entries that
satisfies a characteristic identity, the Filippov identity.
These {\it $n$-Lie} or {\it Filippov algebras} (FA) $\mathfrak{G}$
reduce for $n=2$ to ordinary Lie algebras $\mathfrak{g}$.

The properties of Filippov algebras \cite{Filippov} have been
studied further in parallel with those of the Lie algebras,
specially by Kasymov \cite{Kasymov:88, Kasymov:95} and Ling
\cite{Ling93} (see \cite{review} for a review).
It has been shown, for instance, that it is possible
to define solvable ideals, simple and semisimple Filippov algebras,
etc. Semisimple FAs satisfy a Cartan-like criterion
\cite{Kasymov:95} and, as in the Lie algebra case, they are given by
the direct sums of simple ones. One result, however, in which FAs
differ significantly from their $n=2$ Lie algebra counterparts is
that for each $n>2$ there is only one complex simple finite Filippov
algebra \cite{Filippov,Ling93}, which is $(n+1)$-dimensional. The
real Euclidean simple $n$-Lie algebras $A_{n+1}$, which are
constructed on Euclidean $(n+1)$-dimensional vector spaces, are
thus the only $(n>2)$-Lie (Filippov) algebra generalizations
of the simple $so(3)$ Lie algebra. Similarly, the simple
pseudoeuclidean ones may be considered as $n > 2$ generalizations
of $so(1,2)$.

Other properties of FAs, such as deformations (or {\it e.g.},
central extensions) may be studied. As in the general and Lie
algebra cases \cite{Gers:63,Nij-Rich:67}, deformations are
associated with FA cohomology. The Filippov algebra cohomology
suitable for deformations of Filippov algebras was given in
\cite{Gau:96} in the context of Nambu-Poisson algebras (see further
\cite{Da-Tak:97,Rot:05,Az-Iz:09}); the FA cohomology generalizes the
Lie algebra cohomology complexes (see also \cite{review}). The FA
cohomology is not completely straightforward. For instance, for
$n>3$ it turns out that the $p$-cochains are mappings $\alpha^p:
\wedge^{n-1} \mathfrak{G} \otimes \mathop{\cdots}\limits^{p} \otimes
\wedge^{n-1} \mathfrak{G} \wedge \mathfrak{G} \rightarrow
\mathbb{R}$ ({\it e.g.} in the cohomology suitable for central
extensions of FAs), rather than
 $\alpha^p:\wedge^p\mathfrak{g}\rightarrow \mathbb{R}$
as they would be for Lie algebras $\mathfrak{g}$.
Thus, it is convenient to label the $p$-cochains
by the number $p$ of arguments $\mathcal{X} \in \wedge^{n-1}
\mathfrak{G}$ that they contain rather than by the number of
elements of $\mathfrak{G}$ itself (the $\mathcal{X}$s were called
{\it fundamental objects} in \cite{Az-Iz:09}). It has been proved
recently \cite{Az-Iz:09} that there is a Whitehead lemma for
Filippov algebras: semisimple FAs do not have non-trivial central
extensions and are moreover rigid {\it i.e.}, they do not admit
non-trivial deformations. As a result, the Whitehead lemma holds
true for all $n$-Lie semisimple FAs, $n\ge 2$.

Besides the above finite-dimensional simple FAs there are also
infinite-dimensional simple ones (see \cite{Ca-Kac:10}), as those
defined by the $n$-bracket bracket given by the Jacobian of
functions. This bracket, which satisfies \cite{Filippov,
Filippov:98} the Filippov identity and therefore determines an infinite-dimensional
FA, had actually been considered long before by Nambu
\cite{Nambu:73}. He studied specially the $n=3$ case, as a
generalization of the two-entries Poisson bracket, in an attempt to
introducing a new type of dynamics beyond the standard
Hamilton-Poisson one; the Nambu bracket satisfies additionally
Leibniz's rule. Nambu did not write the Filippov identity that is satisfied by his
bracket; this was done later by Sahoo and Valsakumar
\cite{Sah-Val:92-93} who considerd it as a consistency condition for
the time evolution of Nambu mechanics, as reflected by the
derivation property that is expressed by the Filippov identity. The general $n>3$
case was studied in detail by Takhtajan \cite{Tak:93}, leading to an
$n$-ary generalization of the Poisson structures that he called {\it
Nambu-Poisson structures}. This sparkled an extensive analysis of
various issues related with them, including the notoriously difficult
problem of the quantization of Nambu-Poisson mechanics that
also had been discussed by Nambu himself \cite{Nambu:73}
(and which, in our view, does not admit a completely satisfactory solution,
see \cite{AIPB:97,review}). In the last few years, FAs have reappeared
in physics in another context, namely in the
Bagger-Lambert-Gustavsson model \cite{Ba-La:06-07,Gus:07,Gustav:08},
originally proposed as a candidate for the low-energy effective
action of a system of coincident membranes in M-theory. These and
other physical aspects of FAs are reviewed in \cite{review}, to
which we refer for further information and references.

 In this paper, however, we address a mathematical problem: the
\.In\"on\"u-Wigner type contractions of Filippov algebras. These are
introduced and discussed in generality here. As is well known, all
Filippov algebras $\mathfrak{G}$ have an associated {\it Lie}
algebra Lie$\,\mathfrak{G}$, the algebra of the inner derivations of
$\mathfrak{G}$. Thus, a natural question to ask is whether there is
any relation between the Lie algebra $\mathrm{Lie}\,\mathfrak{G}_c$
associated with some contraction $\mathfrak{G}_c$ of a FA
$\mathfrak{G}$ and a (\.In\"on\"u-Wigner \cite{IW53} (IW) or a
generalized Weimar-Woods \cite{Wei:00} (W-W)) contraction of the Lie
algebra Lie$\,\mathfrak{G}_c$ associated with the contracted FA
$\mathfrak{G}_c$. Clearly $\mathrm{Lie}\,\mathfrak{G}_c \neq
(\mathrm{Lie}\,\mathfrak{G})_c$ in general, but it is still possible
to compare the structure of $(\mathrm{Lie}\,\mathfrak{G})_c$ and
$\mathrm{Lie}\,\mathfrak{G}_c$ for a given $\mathfrak{G}$. We shall
use the simple $A_{n+1}$ FAs to illustrate this point.

The plan of the paper is as follows. Sec.~2 briefly describes the FA
structure, including the fundamental objects $\mathcal{X}$ and the
simple finite-dimensional FAs. Sec.~\ref{sec.LieG} contains the
description of the Lie algebra associated with a FA and, in
particular, considers the case of Lie$\,A_{n+1}=so(n+1)$.
Sec.~\ref{sec.contFAs} is devoted to the description of contractions
$\mathfrak{G}_c$ of arbitrary FAs $\mathfrak{G}$, starting with the
simplest $n=3$ case. Sec.~\ref{sec.LieGc} describes the structure of
the Lie algebra $\mathrm{Lie}\,\mathfrak{G}_c$ associated with a
given contraction $\mathfrak{G}_c$;  Sec.~\ref{Sec.An+1c}
considers the non-trivial contractions $(A_{n+1})_c$ of the
simple $A_{n+1}$ FAs and gives the structure of their associated
Lie$\,(A_{n+1})_c$ Lie algebras. Sec.~\ref{sec.contLie3} discusses
the relation between $\mathrm{Lie}\,\mathfrak{G}_c$ and
$(\mathrm{Lie}\,\mathfrak{G})_c$. To this end, we find the IW and
W-W contractions of $\mathrm{Lie}\,\mathfrak{G}$ for the simple FAs
$A_{n+1}$ that follow the patterns suggested by the structure of
$\mathrm{Lie}\,\mathfrak{G}_c$, and then compare the results with
it. Finally, Sec.~\ref{conclusions} contains some conclusions.

All FAs considered below are real and finite-dimensional.

\section{$n$-Lie or Filippov algebras}
A {\it Filippov algebra} (FA) \cite{Filippov,Kasymov:88} or {\it
$n$-Lie algebra} $\mathfrak{G}$ (see also
\cite{Kasymov:95,Filippov:98,Ling93} and {\it e.g.} \cite{review}
for a review and further references) is a vector space endowed with
a $n$-linear fully skewsymmetric map
$[\,,\,,\,\mathop{\cdots}\limits^{n} \,,\,,\,]:\mathfrak{G} \times
\mathop{\dots}\limits^{n} \times \mathfrak{G} \rightarrow
\mathfrak{G}$ such that the Filippov identity (FI),
\begin{eqnarray}
\label{nFI}
&& [X_1,\ldots,X_{n-1} ,[Y_1,\ldots,Y_n]] =
[[X_1,\ldots,X_{n-1} ,Y_1],Y_2,\ldots , Y_n] +
\\ &&  + [Y_1,[X_1,\ldots,X_{n-1} ,Y_2],Y_3,\ldots, Y_n]
+\ldots +[Y_1 ,\ldots ,Y_{n-1},[X_1,\ldots,X_{n-1},Y_n]]
\nonumber
\end{eqnarray}
is satisfied $\forall X,Y\in \mathfrak{G}$ or, equivalently,
\cite{Gra-Nil-Pet:08,Gustav:08,review}
\begin{equation}
\label{nFIalt}
[[X_{[k_1 }, X_{k_2}, \ldots, X_{k_n}], X_{l_1]}, \ldots ,
X_{l_{n-1}} ]=0 \; ,
\end{equation}
for the elements of a basis $\{X_i\}$ of  $\mathfrak{G}$.
Both the vector space and the FA structure will be
denoted by the same symbol $\mathfrak{G}$; its meaning will be clear
from the context. For $n=2$ the FI becomes the Jacobi identity (JI)
and the Filippov algebra $\mathfrak{G}$ is an ordinary Lie algebra
$\mathfrak{g}$.

\subsection{Structure constants of $n$-Lie algebras}
Once a basis $\{X_l\}$ of $\mathfrak{G}$ is chosen,
the FA bracket may be defined by the $n$-Lie algebra structure constants,
\begin{equation}\label{nFIdef}
[X_{l_1}, \ldots, X_{l_n}]= f_{l_1 \ldots l_n}{}^k X_k \quad ,
\quad l,k=1,\ldots, \textrm{dim}\mathfrak{G} \; .
\end{equation}
The $f_{l_1 \ldots l_n}{}^k$ are fully skewsymmetric in the $l_1 \ldots l_n$ indices
and satisfy the condition
\begin{equation}\label{nFI-str}
f_{k_1\ldots k_n}{}^{l} f_{l_1\dots l_{n-1}l}{}^k =\sum_{i=1}^n
f_{l_1\ldots l_{n-1} k_i}{}^{l} f_{k_1\ldots k_{i-1}l k_{i+1}\ldots
k_n}{}^k \; ,
\end{equation}
which expresses the FI \eqref{nFI} in terms of the structure constants of
$\mathfrak{G}$. The form \eqref{nFIalt} of the FI leads in coordinates to
the expression\footnote{Eqs.~\eqref{nFIalt}, \eqref{FIcoo} are to be compared with the
{\it generalized Jacobi identity (GJI)}
\begin{equation*}
[X_{[l_1 }, \ldots, X_{l_{n-1}}, [X_{k_1}, \ldots , X_{k_{n]}} ]]=0 \quad , \quad
C_{[k_1 \ldots k_n}{}^l C_{l_1 l_2 \ldots l_{n-1}] l }{}^k =0 \quad ,
\end{equation*}
$n$ even, which is the characteristic identity that satisfies
another $n$-ary generalization of Lie algebras, the {\it
generalized} or {\it higher order Lie algebras} \cite{JPA,CMP,HW95},
which will not be considered here (see \cite{review} for a parallel
analysis of Filippov and higher order Lie algebras and their
associated $n$-ary Poisson structures).}
\begin{equation}\label{FIcoo}
f_{[k_1 \ldots k_n}{}^l f_{l_1] l_2 \ldots l_{n-1} l }{}^k =0 \quad .
\end{equation}

\subsection{Fundamental objects of a FA and their properties}
In an $n$-Lie algebra $\mathfrak{G}$ it is convenient to introduce
objects $\mathcal{X}=(X_1,\ldots,X_{n-1}),\, X_i  \in \mathfrak{G}$,
antisymmetric in its $(n-1)$-arguments, $\mathcal{X}\in \wedge^{n-1}(\mathfrak{G})$;
they define inner derivations
of the FA through the adjoint action. This is defined by
\begin{equation}
ad_{\mathcal{X}}:Z \mapsto ad_{\mathcal{X}}Z  \equiv  \mathcal{X}\cdot Z :=
[X_1,\ldots, X_{n-1}, Z]\; , \quad \forall Z \in \mathfrak{G} \quad .
\end{equation}
In terms of $ad_\mathcal{X}=ad_{(X_1,\ldots,X_{n-1})}$, the FI is written as
\begin{equation} \label{FIad}
ad_\mathcal{X}[Y_{l_1},\ldots, Y_{l_n}] =
\sum_{i=1}^n [Y_{l_1}, \ldots, ad_\mathcal{X} Y_{l_i}, \ldots, Y_{l_n}] \; ,
\quad l=1,\ldots \textrm{dim}\,\mathfrak{G} \; ,
\end{equation}
which expresses that $ad_\mathcal{X}\in \textrm{End}\,\mathfrak{G}$
is an inner derivation of the FA $n$-bracket. For convenience, we
refer to the $\mathcal{X} \in \wedge^{n-1}\mathfrak{G}$ as the {\it
fundamental objects} of the $n$-Lie algebra $\mathfrak{G}$. Since
$ad: \wedge^{n-1}\mathfrak{G} \rightarrow
\textrm{End}\,\mathfrak{G}$ may have a non-trivial kernel, the
correspondence between fundamental objects and inner derivations,
$\mathcal{X}_{a_1  \ldots  a_{n-1}} \mapsto ad_{\mathcal{X}_{a_1
\ldots a_{n-1}}}$, is not injective in general:
 $\mathcal{X} \in \mathrm{ker}\,ad$ when $ad_\mathcal{X}$
is the trivial endomorphism of $\mathfrak{G}$. For instance,
ker$\, ad = \wedge^{n-1} \mathfrak{G}$ and $ad$ is trivial if
$\mathfrak{G}$ is abelian.

The coordinates of the
$(\textrm{dim}\,\mathfrak{G} \times \textrm{dim}\,\mathfrak{G})$-dimensional
matrix $ad_{(X_{l_1},\ldots,X_{l_{n-1}})} \equiv ad_{\mathcal{X}_{l_1 \ldots l_{n-1}}}
\in \textrm{End}\,\mathfrak{G}$
are given by
\begin{equation}
ad_{(X_{l_1},\ldots,X_{l_{n-1}})}{}^l{}_k= f_{l_1 \ldots l_{n-1}k}{}^l \quad ,
\quad ad_{(X_{l_1},\ldots,X_{l_{n-1}})} X_k = [X_{l_1},\ldots,X_{l_{n-1}},X_k] =
f_{l_1 \ldots l_{n-1}k}{}^l X_l \;.
\end{equation}
Then, in terms of the structure constants of the FA, the FI \eqref{FIad} takes the form
\begin{equation}
f_{l_1\,\ldots\,l_n}{}^l ad_{(X_{k_1},\ldots,X_{k_{n-1}})} X_l = (-1)^{n-i}
\sum_{i=1}^n f_{k_1\,\ldots\,k_{n-1} \, l_i}{}^l
ad_{(Y_{l_1},\ldots,Y_{l_{i-1}},Y_{l_{i+1}},\ldots,Y_{l_n})} X_l \; .
\end{equation}

Given two fundamental objects $\mathcal{X}\,,\; \mathcal{Y}$
their composition $ \mathcal{X} \cdot \mathcal{Y} \in \wedge^{n-1} \mathfrak{G}$
is the fundamental object given by the formal sum \cite{Gau:96}
\begin{eqnarray}
\label{comp}
\mathcal{X} \cdot \mathcal{Y} &:=&
\sum_{i=1}^{n-1} (Y_1, \ldots, ad_\mathcal{X} Y_i, \ldots, Y_{n-1}) \nonumber \\
 &=& \sum_{i=1}^{n-1} (Y_1, \ldots, [X_1,\ldots, X_{n-1}, Y_i ] , \ldots, Y_{n-1}) \; ,
\end{eqnarray}
which is the natural extension on $\mathcal{Y}\in \wedge^{n-1}
\mathfrak{G}$ of the action of the adjoint derivative
$ad_\mathcal{X}$ on $\mathfrak{G}$; thus, eq.~\eqref{comp} may be
rewritten as
\begin{equation}
\mathcal{X} \cdot \mathcal{Y}=ad_\mathcal{X} \mathcal{Y}\quad .
\end{equation}
 The composition of fundamental objects is not associative.
In fact, due to the FI, the dot product of fundamental objects
$\mathcal{X}$ of an $n$-Lie algebra $\mathfrak{G}$ satisfies the
relation\footnote{In the case of Lie algebras, $n=2$,
$\,\mathcal{X}$ reduces to a single element $X\in \mathfrak{g}$,
$\,X\cdot Y = [X,Y]$ and, of course, $X \cdot (Y \cdot Z) - Y\cdot(X
\cdot Z) = (X \cdot Y) \cdot Z$ is simply the Jacobi identity,
$[X,[Y,Z]]-[Y,[X,Z]]=[[X,Y],Z]$.}
\begin{equation}\label{XYZ}
\mathcal{X} \cdot (\mathcal{Y} \cdot \mathcal{Z}) -\mathcal{Y}
\cdot (\mathcal{X} \cdot \mathcal{Z})
= (\mathcal{X} \cdot \mathcal{Y}) \cdot \mathcal{Z} \qquad \forall \mathcal{X},\;
\mathcal{Y},\; \mathcal{Z} \in \wedge^{n-1} \mathfrak{G}\quad ,
\end{equation}
and, as a result,
\begin{eqnarray}\label{adxady}
&& \mathcal{X} \cdot (\mathcal{Y} \cdot Z) -\mathcal{Y} \cdot (\mathcal{X} \cdot Z)
= (\mathcal{X} \cdot \mathcal{Y}) \cdot Z \qquad  \hbox{or, equivalently,} \nonumber \\
&& \quad ad_\mathcal{X} ad_\mathcal{Y}\, Z -ad_\mathcal{Y} ad_\mathcal{X}\,  Z
= ad_{\mathcal{X} \cdot \mathcal{Y}} \,Z \quad \forall \mathcal{X},\; \mathcal{Y} \;
\in \wedge^{n-1} \mathfrak{G} \; , \; \forall Z\in \mathfrak{G} \quad .
\end{eqnarray}
Thus, the FI may be written as
\begin{equation}\label{oeFI}
[ad_{\mathcal{X}}, ad_{\mathcal{Y}}]=ad_{\mathcal{X} \cdot \mathcal{Y}} \quad ;
\end{equation}
clearly, $ad_{(ad_{\mathcal{X}} \mathcal{Y})}=ad_{\mathcal{X} \cdot \mathcal{Y}}\,$.
Note that although in general $\mathcal{X} \cdot \mathcal{Y} \neq - \mathcal{Y} \cdot \mathcal{X}$,
eq.~\eqref{adxady} is $\mathcal{X} \leftrightarrow \mathcal{Y}$
skewsymmetric, $\, ad_{\mathcal{X} \cdot \mathcal{Y}} = - ad_{\mathcal{Y} \cdot \mathcal{X}}$.

\subsection{The simple Euclidean FAs}\label{simpleFA}

The simple, finite, $(n+1)$-dimensional $n$-Lie algebras constructed
over $(n+1)$-dimensional vector spaces were already given in
\cite{Filippov}, and found to be the only simple ones in
\cite{Ling93}. For the purposes of this paper it will be sufficient
to consider the Euclidean $n$-Lie algebras, constructed over
$(n+1)$-dimensional Euclidean spaces. The Euclidean FAs $A_{n+1}$
\cite{Filippov} are given by eq.~\eqref{nFIdef} where
\begin{equation}
\label{simpleFAstr}
f_{l_1 \ldots l_n}{}^k= \epsilon_{l_1 \ldots l_n}{}^k \quad ;
\end{equation}
the pseudoeuclidean FAs are simply obtained by adding appropriate
signs (it will be sufficient for our purposes here to
restrict ourselves to \eqref{simpleFAstr} when dealing with
simple FAs). Lowering the index $k$ with Euclidean metric the structure
constants are given by the fully skewsymmetric tensor of an
Euclidean $(n+1)$-dimensional vector space.

It is not difficult to check that these algebras are indeed simple.
Clearly, $[\mathfrak{G},\ldots,\mathfrak{G}]\neq \{0\}$ (in fact,
$[\mathfrak{G},\ldots,\mathfrak{G}]=\mathfrak{G}$) and they do not
contain any non-trivial ideal (a subspace $I$ of a FA $\mathfrak{G}$
is an ideal \cite{Filippov,Ling93} if $[\mathfrak{G},
\mathop{\dots}\limits^{n-1},\mathfrak{G},I]\subset I$). Further, the
structure constants (\ref{simpleFAstr}) do define a FA since the
FI is satisfied; we present here a short proof. For $n=3$,
$\mathfrak{G}=A_4$, the four terms in the FI
\begin{eqnarray}\label{3FI-2}
&& [X_{l_1},X_{l_2},[Y_{k_1},Y_{k_2},Y_{k_3}]] =
\\ &&  [[X_{l_1},X_{l_2},Y_{k_1}],Y_{k_2},Y_{k_3}]+[Y_{k_1},[X_{l_1},X_{l_2},Y_{k_2}],Y_{k_3}]
+[Y_{k_1},Y_{k_2},[X_{l_1},X_{l_2},Y_{k_3}]]  \quad ; \nonumber
\end{eqnarray}
are all zero unless two $k$ indices are  equal to the two $l$ ones,
($k_1 , k_2) = (l_1 , l_2$), say, in which case is obviously satisfied
since it reduces to
\begin{equation}
[X_{l_1},X_{l_2},[X_{l_1},X_{l_2},Y_{k_3}]] =
[X_{l_1},X_{l_2},[X_{l_1},X_{l_2},Y_{k_3}]]\quad ,
\end{equation}
which are the only terms that survive since
$[X_{l_1},X_{l_2},Y_{l_3}]=\epsilon_{l_1l_2l_3}{}^{l_4}X_{l_4}$.
This argument is easily extended to general $n$. Since there are
$2n-1$ entries in the double $n$-bracket and $n+1$ elements in the
basis $\{X_l\}$ of $A_{n+1}$, at least $n-2$ elements are
necessarily repeated in the double bracket. Thus, since the separate
$(n-1)$ and $n$ entries in each part of the double bracket cannot
have a repeated element due to the skewsymmetry, we see that the two
parts must have at least $n-2$ equal entries,
$[X_{l_1},\ldots,X_{l_{n-2}},
X_{l_{n-1}},[X_{k_1},\ldots,X_{k_{n-2}},X_{k_{n-1}},X_{k_n}]]$ with
$(l_1,\ldots ,l_{n-2})=(k_1,\ldots ,k_{n-2})$, say. If they only share
these $n-2$ entries all the $n+1$ basis elements will be present in
the double bracket, and then the inner $n$-bracket will necessarily
give rise to an element already present as one of the other $n-1$
entries in the outer bracket, giving zero. If they share $n-1$
entries {\it e.g.}, ($k_1, \ldots ,k_{n-1}) = (l_1, \ldots , l_{n-1}$), the
only non-zero terms in the FI \eqref{nFI} or \eqref{nFIalt}
are the two that do not mix the $k_1 \ldots k_{n-1}$ indices with
the $l_1 \ldots l_{n-1}$ ones, which give the trivial identity
\begin{equation}
[X_{l_1},\ldots,X_{l_{n-1}},[X_{l_1},\ldots,X_{l_{n-1}},Y_{k_n}]] =
[X_{l_1},\ldots,X_{l_{n-1}},[X_{l_1},\ldots,X_{l_{n-1}},Y_{k_n}]]\; .
\end{equation}
Thus, the FI is satisfied by the simple $(n+1)$-dimensional FAs.

When $\mathfrak{G}$ is simple, the composition of two fundamental
objects $\mathcal{X}=(X_{k_1},\ldots,X_{k_{n-1}})=
\mathcal{X}_{k_1\dots k_{n-1} }$ and
$\mathcal{Y}=(X_{j_1},\ldots,X_{j_{n-1}})$ is antisymmetric,
$\mathcal{X} \cdot \mathcal{Y} = - \mathcal{Y} \cdot \mathcal{X}$.
To prove this we take again into account the form
\eqref{simpleFAstr} of the structure constants. Indeed, in
\begin{eqnarray}\label{nsimpleXY}
\mathcal{X}_{k_1\,\ldots\,k_{n-1}} \cdot \mathcal{Y}_{j_1\,\ldots\,j_{n-1}}
=\sum_{i=1}^{n-1} (X_{j_1}, \ldots, [X_{k_1},\ldots, X_{k_{n-1}}, X_{j_i}], \ldots, X_{j_{n-1}}) \nonumber \\
= \sum_{i=1}^{n-1} \epsilon_{k_1\,\ldots\,k_{n-1} \, j_i}{}^{l} (X_{j_1}, \ldots, {X}_{j_{i-1}},X_l, {X}_{j_{i+1}}
\ldots, X_{j_{n-1}} )
\end{eqnarray}
the only nonvanishing terms will be those in which $n-2$ of the indices
$k_1\,\ldots\,k_{n-1}$ are equal to $n-2$ of the indices
$j_1\,\ldots\, j_{n-1}$, since there are $n+1$ basis elements
and the indices $j_i$, $l$, in eq.~\eqref{nsimpleXY}
must be different from both $k_1\,\ldots\,k_{n-1}$ and $j_1\,\ldots\,\hat{j_i}\ldots\,j_{n-1}$.
Taking, $j_i=k_i, \; i=1,\ldots,n-2$, we see that
\begin{eqnarray}\label{sFAxy-yx}
\mathcal{X}_{k_1\,\ldots\,k_{n-1}} \cdot \mathcal{Y}_{k_1 \,\ldots\,k_{n-2}\,j_{n-1}}
= \epsilon_{k_1\,\ldots\,k_{n-1}\, j_{n-1}}{}^{l} ( X_{k_1}, \ldots, {X}_{k_{n-2}}, X_{l})
= - \mathcal{Y}_{k_1\, \ldots\,k_{n-2}\,j_{n-1}} \cdot \mathcal{X}_{k_1 \,\ldots\,k_{n-1}} \, ,\;
\end{eqnarray}
as we wanted to prove.

\section{The Lie algebra Lie$\,\mathfrak{G}$ associated to an $n$-Lie algebra $\mathfrak{G}$}\label{sec.LieG}

The inner or adjoint derivations $ad_\mathcal{X}\in \textrm{End}\, \mathfrak{G}$ associated with the
fundamental objects $\mathcal{X}\in \wedge^{n-1} \mathfrak{G}$ determine an ordinary Lie
algebra for the bracket in $\textrm{End}\, \mathfrak{G}$,
\begin{equation}
\label{Lie-from-FA}
ad_\mathcal{X} ad_\mathcal{Y}  -ad_\mathcal{Y} ad_\mathcal{X}= [ad_\mathcal{X}, ad_\mathcal{Y}] =
ad_{(\mathcal{X} \cdot \mathcal{Y})} \quad.
\end{equation}
Indeed, they satisfy the JI since, using eq.~\eqref{XYZ} and that
$\, ad_{\mathcal{X} \cdot \mathcal{Y}} = - ad_{\mathcal{Y} \cdot \mathcal{X}}$,
\begin{eqnarray}
&& [ad_\mathcal{X},[ad_\mathcal{Y},ad_\mathcal{Z}]]+[ad_\mathcal{Y},
[ad_\mathcal{Z},ad_\mathcal{X}]]+ [ad_\mathcal{Z},[ad_\mathcal{X},ad_\mathcal{Y}]] \nonumber \\
&& =ad_{\mathcal{X} \cdot (\mathcal{Y} \cdot \mathcal{Z})}
+ ad_{\mathcal{Y} \cdot (\mathcal{Z} \cdot \mathcal{X})}
+ ad_{ \mathcal{Z} \cdot (\mathcal{X} \cdot \mathcal{Y}) }
= ad_{\mathcal{X} \cdot (\mathcal{Y} \cdot \mathcal{Z})
- \mathcal{Y} \cdot (\mathcal{X} \cdot \mathcal{Z})
- (\mathcal{X} \cdot \mathcal{Y})\cdot \mathcal{Z} } = 0 \quad \quad.  \label{JIprove}
\end{eqnarray}
This is the Lie algebra Lie$\,\mathfrak{G} \equiv \textrm{InDer} \,
\mathfrak{G} \subset \mathrm{End}\,\mathfrak{G}$ of
inner derivations associated with the FA $\mathfrak{G}$. Clearly,
dim$\,\mathrm{Lie}\,\mathfrak{G}=\left(
                                        \begin{array}{c}
                                        \mathrm{dim}\,\mathfrak{G} \\
                                               n-1 \\
                                                     \end{array}
                                  \right)
-\mathrm{dim}\,(\mathrm{ker}\,ad)$.

If $\mathfrak{G}$ is the simple FA $A_{n+1}$, all derivations are inner;
further, Lie~$A_{n+1}=so(n+1)$
(see {\it e.g.} \cite{Ling93,review})
and, of course,
$\left(
                        \begin{array}{c}
                        n+1 \\
                        n-1 \\
                        \end{array}
                 \right)=\mathrm{dim}\,so(n+1)$.

\subsection{Structure constants of Lie$\,\mathfrak{G}$ for a $3$-Lie algebra}\label{3lie}

For $n=3$ the coordinates of the $\textrm{dim}\,\mathfrak{G} \times \textrm{dim}\,\mathfrak{G}$-dimensional
matrix $[X_{l_1},X_{l_2},\quad] \equiv ad_{(X_{l_1},X_{l_{2}})} \in \textrm{End}\,\mathfrak{G}$
are given by
\begin{equation}\label{3EndG}
ad_{(X_{l_1},X_{l_2})}{}^l{}_k= f_{l_1 l_{2}k}{}^l \quad ,
\quad ad_{(X_{l_1},X_{l_{2}})} X_k = [X_{l_1},X_{l_{2}},X_k] = f_{l_1 l_{2}k}{}^l X_l \;.
\end{equation}
Then, the form \eqref{oeFI} of the FI for $n=3$
$$ad_{(X_{l_1},X_{l_2})}(ad_{(Y_{k_1},Y_{k_2})})
- ad_{(Y_{k_1},Y_{k_2})}(ad_{(X_{l_1},X_{l_2})})
=a d_{([X_{l_1},X_{l_2},Y_{k_1}],Y_{k_2})+(Y_{k_1},[X_{l_1},X_{l_2},Y_{k_2}])}$$
can be written as
\begin{equation}\label{3str-rel1}
[ad_{(X_{l_1},X_{l_2})},ad_{(Y_{k_1},Y_{k_2})}]{}^l{}_k
=f_{l_1 l_2 k_1}{}^j f_{j k_2 l}{}^k + f_{l_1 l_2 k_2}{}^j f_{k_1 j l}{}^k
= - f_{l_1 l_2 [k_1}{}^j f_{k_2] j l}{}^k \;.
\end{equation}
This shows antisymmetry under the interchange of the indices $(k_1 k_2)$ and
$(l_1 l_2)$, {\it i.e.},
$$f_{l_1 l_2 [k_1}{}^j f_{k_2] j l}{}^k = - f_{k_1 k_2 [l_1}{}^j f_{l_2] j l}{}^k\; ,$$
which also follows directly from the
FI $\,f_{[k_1 k_2 l_1}{}^j f_{l_2] l j }{}^k =0\, $ (eq.~\eqref{FIcoo}).

Using eq.~\eqref{3EndG} we can write
\begin{equation}
f_{l_1 l_2 [k_1}{}^j ad_{(X_{k_2]}, X_{j})} = -  f_{k_1 k_2
[l_1}{}^j ad_{(X_{ l_2 ]}, X_{j})} \quad ,
\end{equation}
or, equivalently,
\begin{eqnarray}\label{kkll}
(f_{l_1 l_2 [k_1}{}^j \delta_{k_2]}^l +
f_{k_1 k_2 [l_1}{}^j \delta_{l_2]}^l) ad_{(X_{l}, X_{j})} = 0 \; .
\end{eqnarray}

\noindent Using eq.~\eqref{3str-rel1}, the commutators of Lie$\,\mathfrak{G}$ can be expressed as
\begin{equation}\label{3str-rel2}
[ad_{(X_{l_1},X_{l_2})},ad_{(Y_{k_1},Y_{k_2})}]{}^l{}_k
=\frac12 C_{l_1 l_2 k_1 k_2}{}^{j_1 j_2} ad_{(X_{j_1},X_{j_2})}{}^l{}_k \quad, \quad
C_{l_1 l_2 k_1 k_2}{}^{j_1 j_2} = f_{l_1 l_2 [k_1}{}^{[j_1} \delta_{k_2]}^{j_2]} \;.
\end{equation}

However, this does not mean (see also \cite{Gustav:08}) that the
above $C$'s are the structure constants of Lie$\,\mathfrak{G}$.
Although the {\it r.h.s.} of eq.~\eqref{3str-rel1} is $(l_1 l_2)
\leftrightarrow (k_1 k_2)$ skewsymmetric as mandated by the {\it
l.h.s.}, this does not necessarily imply that the constants $C_{l_1
l_2 k_1 k_2}{}^{j_1 j_2}$ in eq.~\eqref{3str-rel2} retain this
property once the sum over $(j_1 j_2)$ is removed. One may, of
course, write antisymmetric $C$'s in eq.~\eqref{3str-rel2} by taking
\begin{equation}
C_{l_1 l_2 k_1 k_2}{}^{j_1 j_2} =
\frac12 \left( f_{l_1 l_2 [k_1}{}^{[j_1} \delta_{k_2]}^{j_2]} -(l \leftrightarrow k) \right) \; ,
\end{equation}
but this is not sufficient to look at them as structure constants of
Lie$\,\mathfrak{G}$ since, in general, the indices $(j_1 j_2)$ that
characterize $\mathcal{X}_{j_1 j_2}=(X_{j_1}, X_{j_2})$ are not
suitable to label the matrices $ad_{(X_{j_1}, X_{j_2})}$. Since
$\mathcal{X}_{k_1 k_2} \neq \mathcal{X}_{l_1 l_2} \nRightarrow
ad_{\mathcal{X}_{k_1 k_2}} \neq ad_{\mathcal{X}_{l_1 l_2}}$ in
general, the $(j_1,j_2)$-labelled $ad_{(X_{j_1}, X_{j_2})}$ may not
be a basis of Lie$\,\mathfrak{G}$.

The Jacobi identity is of course satisfied by the endomorphisms
$ad_{(X_{s_1},X_{s_2})}$ of $\mathfrak{G}$:
\begin{eqnarray}
\sum_{cycl.\, (j_1 j_2),(k_1 k_2),(l_1 l_2)}
\left( C_{j_1 j_2 k_1 k_2}{}^{r_1 r_2} C_{l_1 l_2 r_1 r_2}{}^{s_1 s_2} \right)
(ad_{(X_{s_1},X_{s_2})})^l{}_k= 0 \; , \label{JIad} \\
\label{jjkk} \left( C_{j_1 j_2 k_1 k_2}{}^{s_1 s_2} + C_{k_1 k_2 j_1 j_2}{}^{s_1 s_2} \right)
(ad_{(X_{s_1},X_{s_2})})^l{}_k = 0 \;,
\end{eqnarray}
but the $ad_{(X_{s_1},X_{s_2})}$ cannot be removed from eqs.~\eqref{JIad} and \eqref{jjkk}. \\

Nevertheless,
\begin{equation}\label{3JI-}
\sum_{cycl.\, (j_1 j_2),(k_1 k_2),(l_1 l_2)}
(C_{j_1 j_2 k_1 k_2}{}^{r_1 r_2} C_{l_1 l_2 r_1 r_2}{}^{s_1 s_2}) =0
\end{equation}
(cf. \eqref{JIad}) holds if the structure constants $C$ in
eq.~\eqref{3str-rel2} are already skewsymmetric under the
interchange $l_1 l_2 \leftrightarrow k_1 k_2$ {\it i.e.}, when
\begin{equation}\label{3simpl}
f_{k_1 k_2 [l_1}{}^{[j_1} \delta_{l_2]}^{j_2]}
= - f_{l_1 l_2 [k_1}{}^{[j_1} \delta_{k_2]}^{j_2]}\;.
\end{equation}
We will see below that this is the case for simple $n$-Lie algebras, for which
{\it e.g.} eq.~\eqref{simpleFAstr} holds, $ad$ is injective and the matrices
$ad_{(X_{j_1}, X_{j_2})}$ define a basis of the associated Lie algebra. For instance, when
$n=3$ it is easy to see that for $A_4$
\begin{equation}\label{3simple}
\epsilon_{k_1 k_2 [l_1}{}^{[j_1} \delta_{l_2]}^{j_2]}
= - \epsilon_{l_1 l_2 [k_1}{}^{[j_1} \delta_{k_2]}^{j_2]}\;,
\end{equation}
since for $\epsilon_{k_1 k_2 [l_1}{}^{[j_1} \delta_{l_2]}^{j_2]}$
to be different from zero we need that one of the indices $k_1$, $k_2$ is equal to
$l_1$ or $l_2$, say $k_2=l_2$, and then
$\epsilon_{k_1 l_2 [l_1}{}^{[j_1} \delta_{l_2]}^{j_2]}
 = - \epsilon_{l_1 l_2 [k_1}{}^{[j_1} \delta_{l_2]}^{j_2]}$ by the
 antisymmetry of the elements in $\epsilon$.
Then, using the relation \eqref{3simple} and the FI \eqref{nFI-str}
for $n=3$, the JI in eq.~\eqref{3JI-} follows.

\subsection{The general $n$-Lie case}\label{c:nlie}

Let now $\mathfrak{G}$ be an $n$-Lie algebra, and $ad_{(X_{k_1},
\ldots, X_{k_{n-1}})}$ the inner derivations associated with the
fundamental objects $\mathcal{X}_{k_1\dots k_{n-1}}\,$,
$$ad_{(X_{k_1}, \ldots, X_{k_{n-1}})} :
Z \rightarrow [X_{k_1}, \ldots, X_{k_{n-1}},Z] \in \mathfrak{G}\;.$$
The $ad_\mathcal{X}$ determine the Lie algebra Lie$\,\mathfrak{G}$
associated with the FA $\mathfrak{G}$. In terms of components, the
commutators of the elements $ad_{\mathcal{X}}\in$
Lie$\,\mathfrak{G}$ can be written as:
\begin{eqnarray}
\label{nLieG}
[ad_{\mathcal{X}}, ad_{\mathcal{Y}}]
 &=& [ad_{(X_{k_1}, \ldots, X_{k_{n-1}})}, ad_{(X_{j_1}, \ldots, X_{j_{n-1}})}]
 = \frac12 ad_{(\mathcal{X} \cdot \mathcal{Y} - \mathcal{Y} \cdot \mathcal{X})}
 = \nonumber \\
 &=& \frac12 \sum_{i=1}^{n-1} \left(f_{{k_1}\,\ldots\,
 {k_{n-1}}\, {j_i}}{}^{l} ad_{(X_{j_1} , \ldots,
    {X}_{j_{i-1}}, X_l, {X}_{j_{i+1}} \ldots, X_{j_{n-1}})} \right. \nonumber \\
 && \left. - f_{{j_1}\,\ldots\, {j_{n-1}}\, {k_i}}{}^{l}
 ad_{(X_{k_1}, \ldots, {X}_{k_{i-1}}, X_l, {X}_{k_{i+1}} \ldots, X_{k_{n-1}})}\right)
 \nonumber \\&\equiv &
 \frac{1}{(n-1)!}C_{k_1\,\ldots\,k_{n-1}\, j_1\,\ldots \, j_{n-1}}{}^{l_1 \, \ldots \, l_{n-1}}
    ad_{(X_{l_1},\ldots, X_{l_{n-1}})} \; ,
\end{eqnarray}
where we have taken
\begin{equation}\label{str-cont}
C_{k_1\,\ldots\,k_{n-1} \,j_1\,\ldots\, j_{n-1}}{}^{l_1 \, \ldots \, l_{n-1}} =
\frac{1}{2(n-2)!} \left( f_{{k_1}\,\ldots\, {k_{n-1}}\, {[j_1}}{}^{[l_1}\delta_{j_2}^{l_2} \ldots
\delta_{j_{n-1}]}^{l_{n-1}]} - (k\leftrightarrow j) \right) \quad ,
\end{equation}
so that they are antisymmetric under the permutation of the indices
$(k_1,\ldots,k_{n-1})$ and $(j_1,\ldots,j_{n-1})$.

The Jacobi identity for Lie$\,\mathfrak{G}$ (cf. eq.~\eqref{JIad}) reads
\begin{eqnarray}\label{JInLieG}
\sum_{cycl. \; j,\,k,\,l} C_{j_1\,\ldots\,j_{n-1} \,k_1\,\ldots\,
k_{n-1}}{}^{h_1 \, \ldots \, h_{n-1}} C_{l_1\,\ldots\,l_{n-1}
\,h_1\,\ldots\, h_{n-1}}{}^{i_1 \, \ldots \, i_{n-1}} ad_{(X_{i_1},
\ldots, {X}_{i_{n-1}})} =0 \;.
\end{eqnarray}
As in the $n=3$ case, it is possible to remove the $ad_\mathcal{X}$ above when
$\{ad_{(X_{i_1}, \ldots, {X}_{i_{n-1}})}\}$ is a basis of Lie$\,\mathfrak{G}$,
{\it i.e.}, when $ad$ is injective. This is the case for the simple FAs, for which the
terms $f_{{k_1}\,\ldots\, {k_{n-1}}\, {[j_1}}{}^{[l_1}\delta_{j_2}^{l_2} \ldots
\delta_{j_{n-1}]}^{l_{n-1}]}$ are skewsymmetric under the interchange
$(k_1,\ldots,k_{n-1}) \leftrightarrow (j_1,\ldots,j_{n-1})$.
The proof is familiar by now (see Sec.~\ref{simpleFA}):
the only non-vanishing structure constants of Lie$\,\mathfrak{G}$
for a simple FA are of the form
$C_{k_1\,\ldots\,k_{n-1} \,j_1\,\ldots\, j_{n-1}}{}^{l_1 \, \ldots \, l_{n-1}}$
with $n-2$ of the indices $k_1\,\ldots\,k_{n-1}$ equal to $n-2$ of the indices
$j_1\,\ldots\, j_{n-1}$. Taking again $k_i=j_i, \, i=1,\ldots,n-2$, it follows that
\begin{eqnarray} \label{skewsymmetry}
C_{k_1\,\ldots\,k_{n-2}\,k_{n-1} \,k_1\,\ldots\,k_{n-2}\,j_{n-1}}{}^{l_1 \, \ldots \, l_{n-1}} =
\frac{1}{(n-2)!2} \left(-\epsilon_{k_1\,\ldots\,k_{n-1}\,
[j_{n-1}}{}^{[l_1}\delta_{k_2}^{l_2} \ldots \delta_{k_{n-2}}^{l_{n-2}}
\delta_{k_{1}]}^{l_{n-1}]} + (k \leftrightarrow j)\right) = \nonumber\\
\frac{1}{(n-2)!} \epsilon_{k_1\,\ldots\,k_{n-2}\,j_{n-1}\,
[k_{n-1}}{}^{[l_1}\delta_{k_2}^{l_2} \ldots \delta_{k_{n-2}}^{l_{n-2}}
\delta_{k_{1}]}^{l_{n-1}]}=
-C_{k_1\,\ldots\,k_{n-2}\,j_{n-1} \,k_1\,\ldots\,k_{n-2}\,k_{n-1}}{}^{l_1 \, \ldots \, l_{n-1}}  . \qquad
\end{eqnarray}

\subsection{A trivial example: Lie$\,A_4=so(4)$}\label{sec.LieA4}
Since this case will be used later on, consider $A_4$. It is given by
\begin{equation}\label{A4}
[X_{j_1},X_{j_2},X_{j_3}]= \epsilon_{{j_1}\,{j_2}\,{j_3}}{}^{j_4} X_{j_4} \, , \quad j=1,2,3,4 \, .
\end{equation}
\noindent Lie$\,A_4$ is given by the commutators (cf. eq.~\eqref{3str-rel2})
\begin{eqnarray}\label{LieA4}
&&[ad_{(X_{k_1},X_{k_2})}, ad_{(X_{l_1},X_{l_2})}]
= ad_{([X_{k_1}, X_{k_{2}}, X_{l_1}], X_{l_{2}})}
+ ad_{(X_{l_{1}},  [X_{k_1}, X_{k_{2}}, X_{l_2}])} = \nonumber \\
&& \qquad = \epsilon_{k_1\, k_2\, l_1}{}^l ad_{(X_l, X_{l_2})}
+ \epsilon_{k_1\, k_2\, l_2}{}^l ad_{(X_{l_1}, X_{l})}
= \frac12 C_{k_1 \, k_2 \, l_1 \, l_2}{}^{j_1\, j_2} ad_{(X_{j_1},X_{j_2})}  \; ,  \quad
\end{eqnarray}
where the structure constants of Lie$\,A_4$ are given by
\begin{equation}
C_{k_1 \, k_2 \, l_1 \, l_2}{}^{j_1\, j_2}= - C_{l_1 \, l_2 \,k_1 \, k_2 }{}^{j_1\, j_2}
= \epsilon_{k_1\, k_2\, [l_1}{}^{[j_1} \delta^{j_2]}_{l_2]}  \, ;
\end{equation}
they may be non-zero only if one of the indices ${k_1, \,k_2}$ is
equal to one of the indices ${l_1, \,l_2}$, as seen in Sec.~\ref{3lie}.

Let the $\mathfrak{G}$ vector space be split into the space $\mathfrak{G}_0$ generated by one generator,
say $X_4$, and the subspace $\mathfrak{V}$ generated by the remaining elements of the $A_4$ basis,
\begin{equation}
\mathfrak{G}=\mathfrak{G}_0\oplus \mathfrak{V}\quad , \quad \mathfrak{G}_0= \langle X_4 \rangle \quad,
\quad \mathfrak{V}=\langle X_u, \; u=1,2,3\rangle .
\end{equation}
This type of splitting will prove useful when considering the
contractions of $\mathfrak{G}$ since $\mathfrak{G}_0$ is obviously a
subalgebra of $\mathfrak{G}$. To look at Lie$\,A_4$ we split its
vector space into subspaces $\langle ad_{(X_{j_1}, X_{j_2})}
\rangle$ according to the number of elements of $\mathfrak{V}$ that
appear in the fundamental objects in the inner derivations
$ad_{(X_{j_1}, X_{j_2})}$ that generate each of them. Then,
\begin{eqnarray}\label{splittingLieG3}
&& {\mathcal{W}^{(0)}}= \langle ad_{(X_{4}, X_{4})}\rangle  = \{0\} \, ,\nonumber\\
&& {\mathcal{W}^{(1)}}= \langle ad_{(X_{4}, X_{u})}\rangle
= \langle ad_{(X_{4}, X_{1})}, ad_{(X_{4}, X_{2})},ad_{(X_{4}, X_{3})}\rangle \, , \; \nonumber \\
&& {\mathcal{W}^{(2)}}= \langle ad_{(X_{u_1}, X_{u_2})}\rangle
= \langle ad_{(X_{2}, X_{3})},ad_{(X_{3}, X_{1})},ad_{(X_{1}, X_{2})}\rangle \, ,
\end{eqnarray}
where we have included $\mathcal{W}^{(0)}$ separately although here
reduces trivially to the zero element of Lie$\, A_4$. The
commutation relations of Lie$\, A_4$,
\begin{eqnarray}\label{LieA4sp-1}
[ad_{\mathcal{X}_{4\,u_1}^{(1)}}, ad_{\mathcal{Y}_{4\,u_2}^{(1)}}]
&\equiv& [ad_{(X_4,X_{u_1})}, ad_{(X_4, X_{u_2})}]
= \epsilon_{4 \, u_1 \, u_2}{}^u ad_{(X_4,X_u)} \in {\mathcal{W}^{(1)}}  \\
\label{LieA4sp-2}{}[ad_{\mathcal{X}_{4\,u_1}^{(1)}}, ad_{\mathcal{Y}_{u_1\,v_2}^{(2)}}]
&\equiv& [ad_{(X_4,X_{u_1})}, ad_{(X_{u_1}, X_{v_2})}]
= \epsilon_{4 \, u_1 \, v_2}{}^u ad_{(X_{u_1},X_{u})} \in {\mathcal{W}^{(2)}}  \\
\label{LieA4sp-3}{}[ad_{\mathcal{X}_{u_1\,u_2}^{(2)}}, ad_{\mathcal{Y}_{u_1\,v_2}^{(2)}}]
&\equiv& [ad_{(X_{u_1},X_{u_2})}, ad_{(X_{u_1}, X_{v_2})}]
= \epsilon_{u_1 \, u_2 \, v_2}{}^4 ad_{(X_{u_1},X_{4})} \in {\mathcal{W}^{(1)}} \; ,
\end{eqnarray}
show that ${\mathcal{W}^{(1)}}$ is a $so(3)$ subalgebra. Renaming the elements of Lie$\,A_4$ as
\begin{eqnarray}\label{so(3)}
Y_1= ad_{(X_{4}, X_{1})},\; Y_2= ad_{(X_{4}, X_{2})},\; Y_3=ad_{(X_{4}, X_{3})} \, , \; \nonumber \\
Z_1= ad_{(X_{2}, X_{3})},\; Z_2= ad_{(X_{3}, X_{1})},\; Z_3=ad_{(X_{1}, X_{2})} \, ,
\end{eqnarray}
eqs.~\eqref{LieA4sp-1}-\eqref{LieA4sp-3} can be written as
\begin{equation}\label{so(4)}
[Y_i,Y_j]=\epsilon_{ij}{}^k Y_k,\quad [Y_i,Z_j]
=\epsilon_{ij}{}^k Z_k,\quad [Z_i,Z_j]=\epsilon_{ij}{}^k Y_k,\quad i,j,k=1,2,3\; ,
\end{equation}
or, with  $\tilde{Y}_i=\frac{1}{2}(Y_i+Z_i)$, $\tilde{Z}_i=\frac{1}{2}(Y_i-Z_i)$,
\begin{equation}\label{so(4)tilde}
[\tilde{Y}_i,\tilde{Y}_j]=\epsilon_{ij}{}^k \tilde{Y}_k,\quad [\tilde{Y}_i,\tilde{Z}_j]=0,
\quad [\tilde{Z}_i,\tilde{Z}_j]=\epsilon_{ij}{}^k \tilde{Z}_k,\quad  i,j,k=1,2,3 \; .
\end{equation}
In fact, as is well known, Lie$\,A_4=so(4)=so(3)\oplus so(3)$ (Lie$\, A_{n+1}$ is simple
but for $n=3$) and dim$\,\mathrm{Lie} \, A_4=6$.

\section{Contractions of FAs} \label{sec.contFAs}
\subsection{The case of $3$-Lie algebras} \label{sec.3Lie}
As is well known, the \.In\"on\"u-Wigner (IW) contraction
\cite{IW53} of a Lie algebra $\mathfrak{g}$ is performed with
respect to a subalgebra $\mathfrak{g}_0 \subset \mathfrak{g}$ by
rescaling the generators of the coset $\mathfrak{g}/\mathfrak{g}_0$
and then taking the contraction limit for the scaling parameter; this
guarantees that the result is also a Lie algebra, $\mathfrak{g}_c$.
Let $\mathfrak{g}$ be defined by
\begin{equation}
[X_{l_1}, X_{l_2}]=f_{l_1\, l_2}{}^l X_l\, , \quad  X_l\, \in \, \mathfrak{g}\,,
\quad l=1,\ldots \textrm{dim}\,\mathfrak{g} \quad ,
\end{equation}
and split its underlying vector space as the sum
$\mathfrak{g}=\mathfrak{g}_0 \oplus \mathfrak{v}$,
$$\mathfrak{g}_0 = \{X_a,\, a=1,\ldots, \textrm{dim}\,\mathfrak{g}_0\}\,,\quad \mathfrak{v}
= \{X_u,\, u=\textrm{dim}\,\mathfrak{g}_0 +1,\ldots,
\textrm{dim}\,\mathfrak{g}\} \; . $$ Then, redefining the basis of
$\mathfrak{v}$ as $X^\prime_u=\epsilon X_u$ and taking the limit
$\epsilon \rightarrow 0$ the contracted algebra $\mathfrak{g}_c$ is
obtained. The generators of $\mathfrak{v}=\mathfrak{g}/\mathfrak{g}_0$
become abelian in $\mathfrak{g}_c$, the preserved subalgebra
$\mathfrak{g}_0 \subset \mathfrak{g}_c$ acts on them and $\mathfrak{g}_c$
has the semidirect structure
 $\mathfrak{g}_c={\mathfrak{v}} \, {\supset \!\!\!\!\!\!
\raisebox{1.5pt} {\tiny +}} \;\,{\mathfrak{g}_0}$, where
$\mathfrak{v}$ is an abelian ideal of $\mathfrak{g}_c$. Obviously,
the IW contraction is dimension preserving.

To generalize the contraction procedure to $n>2$ Filippov algebras,
consider first the simplest case of a 3-Lie algebra $\mathfrak{G}$
given by
\begin{equation}\label{3brackets}
[X_{l_1}, X_{l_2}, X_{l_3}]=
f_{l_1\, l_2\, l_3 }{}^l X_l\quad , \quad X_l \in \mathfrak{G}, \; l =1,\dots \textrm{dim} \,\mathfrak{G}\, .
\end{equation}
The three-bracket satisfies the FI (eq.~\eqref{FIad}), namely
\begin{equation}
ad_\mathcal{X} [Y_{k_1},Y_{k_2},Y_{k_3}] =
[ad_\mathcal{X}Y_{k_1},Y_{k_2},Y_{k_3}]+
[Y_{k_1},ad_\mathcal{X} Y_{k_2}, Y_{k_3}]+[Y_{k_1},Y_{k_2},ad_\mathcal{X} Y_{k_3}]\;.
\end{equation}

\noindent Let $\mathfrak{G}_0\subset \mathfrak{G}$ be a Filippov subalgebra,
$[\mathfrak{G}_0, \mathfrak{G}_0, \mathfrak{G}_0] \subset \mathfrak{G}_0$ and let us split
the $\mathfrak{G}$ vector space as
\begin{equation}
\label{3splitting}
\mathfrak{G}=\mathfrak{G}_0 \oplus \mathfrak{V}\; ,\quad
\begin{array}{l}
\mathfrak{G}_0=\langle X_{a}\rangle \,,  \; a \in I_0 =
\{1,\ldots, \textrm{dim} \mathfrak{G}_0\} \quad \,\\
\mathfrak{V}\;=\langle X_{u}\rangle \,,  \; u \in I_1 =
\{\textrm{dim} \mathfrak{G}_0 +1,\ldots, \textrm{dim} \mathfrak{G}\} \quad ,\\
\end{array}
\end{equation}
where the indices $a$, $b$, $c\,$ label the elements
of a basis of $\mathfrak{G}_0$, $u$, $v$, $w$ refer to the basis
of $\mathfrak{V}$ and the indices $j$, $k$, $l$ refer to the basis
of the FA $\mathfrak{G}$,
\begin{equation}
\label{notation}
X_a, X_b, X_c \in \mathfrak{G}_0  \;,\quad
X_u, X_v, X_w \in \mathfrak{V} \;,
\quad X_j, X_k, X_l \in \mathfrak{G} \,.
\end{equation}

We now define the contraction with respect to the Filippov subalgebra
$\mathfrak{G}_0$ by rescaling the basis elements of
$\mathfrak{V}$, $X^\prime_a=X_a$, $X_u^\prime =
\epsilon X_u$. The four types of brackets in the new, primed basis of $\mathfrak{G}$ are
\begin{eqnarray}
\label{condsubal}
&& [X^\prime_{a_1},X^\prime_{a_2},X^\prime_{a_3}] =
f_{a_1\,a_2\,a_3}{}^a X^\prime_a  \quad +
\underbrace{\epsilon^{-1} f_{a_1\,a_2\,a_3}{}^u X^\prime_u}_{=0\;
(\mathfrak{G}_0\; \mathrm{subalgebra})} \; , \\
&& [X^\prime_{a_1},X^\prime_{a_2},X^\prime_{u_1}]=
\epsilon f_{a_1\,a_2\,u_1}{}^a X^\prime_a + f_{a_1\,a_2\,u_1}{}^u X^\prime_u \; ,\\
&& [X^\prime_{a_1},X^\prime_{u_1},X^\prime_{u_2}]=
\epsilon^2 f_{a_1\,u_1\,u_2}{}^a X^\prime_a + \epsilon f_{a_1\,u_1\,u_2}{}^u X^\prime_u \; ,  \\
&& [X^\prime_{u_1},X^\prime_{u_2},X^\prime_{u_3}]=
\epsilon^3 f_{u_1\,u_2\,u_3}{}^a X^\prime_a + \epsilon^2 f_{u_1\,u_2\,u_3}{}^u X^\prime_u \; .
\end{eqnarray}
Thus, to be able to take the $\epsilon \rightarrow 0$ contraction
limit it is required that $f_{a_1\,a_2\,a_3}{}^u=0$  in
\eqref{condsubal} {\it i.e.}, $\mathfrak{G}_0$ must be a subalgebra
of $\mathfrak{G}$ as originally assumed. Then, the {\it contracted
3-Lie algebra} $\mathfrak{G}_c$,
\begin{equation}\label{3commutators}
[X^\prime_{l_1}, X^\prime_{l_2} , X^\prime_{l_3}]= f^{\prime}_{l_1\, l_2 \, l_3}{}^l X^\prime_l\, ,
\end{equation}
is defined by the FA structure constants
\begin{eqnarray}\label{3str}
f^{\prime}_{l_1\, l_2 \, l_3}{}^l =
\left\{
\begin{array}{ll}
f^\prime_{a_1 \, a_2 \, a_3}{}^a = f_{a_1 \, a_2 \, a_3}{}^a \;,
&  a_1,\, a_2,\, a_3,\, a \, \in \, I_0 \\
f^\prime_{a_1 \, a_2 \, a_3}{}^u = 0 \; ,
& a_1,\, a_2,\, a_3 \, \in I_0 \, ; \;\; u \, \in \, I_1 \\
f^\prime_{a_1 \, a_2 \, u_3}{}^a = 0  \;,
& a_1,\, a_2,\, a \, \in \, I_0 \,; \;\; u_3 \, \in I_1   \\
f^\prime_{a_1 \, a_2 \, u_3}{}^u = f_{a_1 \, a_2 \, u_3}{}^u \; ,
& a_1,\, a_2\, \in I_0 \, ; \;\; u_3 , \, u \, \in \, I_1 \\
f^\prime_{a_1 \, u_2 \, u_3}{}^l = 0  \;,
& a_1\, \in \, I_0 \, ; \;\; u_2,\, u_3\, \in I_1 \, ;\;\;  l \in I_0 \cup I_1     \\
f^\prime_{u_1 \, u_2 \, u_3}{}^l = 0  \;,
& u_1,\, u_2,\, u_3\, \in I_1 \, ;\;\;  l \in I_0 \cup I_1  \quad ,   \\
\end{array}
\right.
\end{eqnarray}
since the FI is obviously satisfied (this will be shown in general
in Sec.~\ref{Sec.Gc}). The $\mathfrak{G}_0 \subset \mathfrak{G}$ subalgebra
is preserved in the contraction process and
dim$\,\mathfrak{G}_c$=dim$\,\mathfrak{G}$.

Of course, once $\mathfrak{G}_c$ has been obtained the primes may be
removed throughout; we shall keep them nevertheless to indicate that
we refer to the structure constants of the contracted FA.
Eq.~\eqref{3str} shows that $\mathfrak{V}$ becomes a FA ideal in
$\mathfrak{G}_c$, as it is the case for the IW contraction of Lie
algebras. Hence, the general structure of the contracted 3-Lie
algebra $\mathfrak{G}_c$ is
\begin{eqnarray}
\label{3liegcstr-1}
&& \begin{array}{l}
\, [\mathfrak{G}_0, \mathfrak{G}_0, \mathfrak{G}_0] \subset \mathfrak{G}_0 \;
\Rightarrow \mathfrak{G}_0 \; \textrm{subalgebra}
\end{array} \\
\label{3liegcstr-2} && \left.
\begin{array}{l}
\,[\mathfrak{G}_0, \mathfrak{G}_0, \mathfrak{V}] \subset \mathfrak{V} \\
\,[\mathfrak{G}_0, \mathfrak{V}, \mathfrak{V}] =0 \\
\,[\mathfrak{V}, \mathfrak{V}, \mathfrak{V}] =0 \\
\end{array}
\right\}
\Rightarrow \mathfrak{V}\; \textrm{abelian ideal} \quad .
\end{eqnarray}
$\mathfrak{V}$ is an ideal because $[\mathfrak{G}, \mathfrak{G},
\mathfrak{V}] \subset \mathfrak{V}$ and abelian by the last
equality. As a result, {\it the contracted FA has the semidirect
structure}\footnote{ We introduce the {\it semidirect extension of
FAs} in similarity with the Lie algebra case. Let $\mathfrak{G}$ be
an $n$-Lie algebra, $\mathfrak{G}_0$ a subalgebra and let
$\mathfrak{G}=\mathfrak{V}\oplus\mathfrak{G}_0$ as a vector space.
Then, $\mathfrak{G}$ is the semidirect FA extension of
$\mathfrak{G}_0$ by $\mathfrak{V}$, $\mathfrak{G}={\mathfrak{V}} \,
{\supset \!\!\!\!\!\! \raisebox{1.5pt} {\tiny +}}\;\,
{\mathfrak{G}_0}$ if $\mathfrak{V}$ is an ideal of $\mathfrak{G}$
and $\mathfrak{G}_0$ acts on it through the (adjoint) action that
results from $\mathfrak{G}_0$ being a subalgebra of $\mathfrak{G}$.
For $n=2$, this recovers the semidirect sum of Lie algebras.}
$\mathfrak{G}_c={\mathfrak{V}} \, {\supset \!\!\!\!\!\!
\raisebox{1.5pt} {\tiny +}}\;\,{\mathfrak{G}_0}$ since
${\mathfrak{G}_0}\subset \mathfrak{G}_c$ acts on the (abelian) ideal
$\mathfrak{V} \subset \mathfrak{G}_c$ through the adjoint action,
$ad_{\mathcal{X}_0}: \mathfrak{V} \mapsto \mathfrak{V}, \;
\mathcal{X}_0 \in \wedge^2 \mathfrak{G}_0$, by the first expression
in eq.~\eqref{3liegcstr-2}. Of course, for $n=2$ this reproduces the
familiar semidirect structure $\mathfrak{g}={\mathfrak{v}} \,
{\supset \!\!\!\!\!\! \raisebox{1.5pt} {\tiny +}}\;\,
{\mathfrak{g}_0}$ of the IW contraction of Lie algebras.

\subsubsection{The Lie algebra Lie$\,\mathfrak{G}_c$ associated with the
contracted 3-Lie algebra $\mathfrak{G}_c$}
\label{sec.3LieGc}

Let now Lie$\,\mathfrak{G}_c \subset \mathrm{End}\, \mathfrak{G}_c$ be the Lie algebra
associated (Sec.~\ref{sec.LieG}) with the contracted 3-Lie
algebra $\mathfrak{G}_c$.
To study its structure let us split the fundamental objects of
$\mathfrak{G}_c$, $\mathcal{X}^\prime \in \wedge^2 \mathfrak{G}_c$,
into three types, as suggested by the splitting of the elements of
$\mathfrak{G}$ itself in eq.~\eqref{3splitting},
\begin{eqnarray}\label{3liegsplitting}
\left.
\begin{array}{l}
\mathcal{X}^{\prime(0)}= (X^\prime_{a_1}, X^\prime_{a_2}) \,   \\
\mathcal{X}^{\prime(1)}= (X^\prime_{a_1}, X^\prime_{u_2}) \,   \\
\mathcal{X}^{\prime(2)}= (X^\prime_{u_1}, X^\prime_{u_2}) \, \quad   \\
\end{array} \right| \qquad
X^\prime_{a_i} \in \mathfrak{G}_0 \, , \;
X^\prime_{u_i}\in \mathfrak{V} \, , \quad \mathfrak{G}_c=\mathfrak{G}_0 \oplus \mathfrak{V} \; .
\end{eqnarray}
{}From the structure of $\mathfrak{G}_c$ (see  eq.~\eqref{3str}) it follows that
\begin{eqnarray}\label{adxz}
&& ad_{\mathcal{X}^{\prime (0)}}X^\prime_a \in \mathfrak{G}_0  \\
&& ad_{\mathcal{X}^{\prime (0)}}X^\prime_u \in \mathfrak{V}  \\
&& ad_{\mathcal{X}^{\prime (1)}}X^\prime_a \in \mathfrak{V}  \\
&& ad_{\mathcal{X}^{\prime (1)}}X^\prime_u = 0  \\
&& ad_{\mathcal{X}^{\prime (2)}}X^\prime_l= 0 \; .  \label{adxz-5}
\end{eqnarray}
The inner derivations $ad_{(X^\prime_{l_1},X^\prime_{l_2})}$ above do not determine a basis of
Lie$\,\mathfrak{G}_c$ since no contracted FA may be
simple and $ad$ is not injective for $\mathfrak{G}_c$. In particular, by eq.~\eqref{adxz-5},
all the elements in ${\mathcal{X}^{\prime (2)}}$ induce the zero derivation,
$\mathcal{X}^{\prime (2)}\in \mathrm{ker}\,ad $.\\

Let $\mathcal{W}^{\prime (r)}=\langle ad_{\mathcal{X}^{\prime(r)}}\rangle$, $r=0,1,2$, be
the vector spaces generated by the inner endomorphisms associated with the
fundamental objects $\mathcal{X}^{\prime(r)}$ in eq.~\eqref{3liegsplitting} ($\mathcal{W}^{\prime(2)}$
is actually zero by eq.~\eqref{adxz-5}).
Lie$\,\mathfrak{G}_c$ is now readily determined from the structure of
$\mathfrak{G}_c$, eqs.~\eqref{3commutators}, \eqref{3str}.
The different types of Lie$\,\mathfrak{G}_c$ commutators are, explicitly,
\begin{eqnarray}
&& [ad_{\mathcal{X}_{a_1 a_2}^{\prime (0)}}, ad_{\mathcal{Y}_{b_1 b_2}^{\prime (0)}}]
 = \frac12 ad_{[(X^\prime_{a_1}, X^\prime_{a_2}) \cdot (X^\prime_{b_1}, X^\prime_{b_{2}})
 - (X^\prime_{b_1}, X^\prime_{b_2}) \cdot (X^\prime_{a_1}, X^\prime_{a_{2}})]} \nonumber\\
&& \quad =  \frac12 \left( f^\prime_{{a_1}\, {a_2}\, {b_1}}{}^{a} ad_{(X^\prime_{a}, X^\prime_{b_2})}
+ f^\prime_{{a_1}\, {a_2}\, {b_2}}{}^{a} ad_{(X^\prime_{b_1}, X^\prime_a)}
-f^\prime_{{b_1}\, {b_2}\, {a_1}}{}^{a} ad_{(X^\prime_{a}, X^\prime_{a_2})}
- f^\prime_{{b_1}\, {b_2}\, {a_2}}{}^{a} ad_{(X^\prime_{a_1}, X^\prime_a)} \right)\;  \nonumber
\\ && \quad \in \mathcal{W}^{\prime(0)} \qquad \label{gc-1}
\\[0.5cm]
&& [ad_{\mathcal{X}_{a_1a_2}^{\prime (0)}}, ad_{\mathcal{Y}_{b_1u_2}^{\prime (1)}}]  =
\frac12 ad_{[(X^\prime_{a_1}, X^\prime_{a_2}) \cdot (X^\prime_{b_1}, X^\prime_{u_2})
-(X^\prime_{b_1}, X^\prime_{u_2}) \cdot (X^\prime_{a_1}, X^\prime_{a_2})]} \nonumber \\
&& \quad = \frac12 \left( f^\prime_{{a_1}\, {a_2}\, {b_1}}{}^{a} ad_{(X^\prime_{a}, X^\prime_{u_2})}
+ \underbrace{f^\prime_{{a_1}\, {a_2}\, {b_1}}{}^{u}}_{=0} ad_{(X^\prime_{u}, X^\prime_{u_2})}
+ \underbrace{f^\prime_{{a_1}\, {a_2}\, {u_2}}{}^{a}}_{=0} ad_{(X^\prime_{b_1}, X^\prime_a)}
+ f^\prime_{{a_1}\, {a_2}\, {u_2}}{}^{u} ad_{(X^\prime_{b_1}, X^\prime_u)} \right. \nonumber \\
&& \quad  \left.- \underbrace{f^\prime_{{b_1}\, {u_2}\, {a_1}}{}^{a}}_{=0}
ad_{(X^\prime_{a}, X^\prime_{a_2})}
- f^\prime_{{b_1}\, {u_2}\, {a_1}}{}^{u} ad_{(X^\prime_u, X^\prime_{a_2})}
- \underbrace{f^\prime_{{b_1}\, {u_2}\, {a_2}}{}^{a}}_{=0} ad_{(X^\prime_{a_1}, X^\prime_a)}
- f^\prime_{{b_1}\, {u_2}\, {a_2}}{}^{u} ad_{(X^\prime_{a_1}, X^\prime_u)} \right) \nonumber
\\ && \quad \in \mathcal{W}^{\prime(1)} \qquad
 \\[0.5cm]
&& [ad_{\mathcal{X}_{a_1u_1}^{\prime (1)}}, ad_{\mathcal{Y}_{a_2 u_2}^{\prime (1)}}]  =
\frac12 ad_{[(X^\prime_{a_1}, X^\prime_{u_1}) \cdot (X^\prime_{a_2}, X^\prime_{u_2})-
(X^\prime_{a_2}, X^\prime_{u_2}) \cdot (X^\prime_{a_1}, X^\prime_{u_1})]} \nonumber\\
&& \quad = \frac12 \left( \underbrace{f^\prime_{{a_1}\, {u_1}\, {a_2}}{}^{a}}_{=0}
ad_{(X^\prime_{a}, X^\prime_{u_2})}
+ f^\prime_{{a_1}\, {u_1}\, {a_2}}{}^{u} \underbrace{ad_{(X^\prime_{u}, X^\prime_{u_2})}}_{=0}
+ \underbrace{f^\prime_{{a_1}\, {u_1}\, {u_2}}{}^{l}}_{=\,0}
ad_{(X^\prime_{a_2}, X^\prime_l)} \right.\nonumber \\
&& \left. \quad -  \underbrace{f^\prime_{{a_2}\, {u_2}\, {a_1}}{}^{a}}_{=0}
ad_{(X^\prime_{a}, X^\prime_{u_1})}
- f^\prime_{{a_2}\, {u_2}\, {a_1}}{}^{u} \underbrace{ad_{(X^\prime_{u}, X^\prime_{u_1})}}_{=0}
- \underbrace{f^\prime_{{a_2}\, {u_2}\, {u_1}}{}^{l}}_{=\,0}
ad_{(X^\prime_{a_1}, X^\prime_l)}\right) =0 \label{gc-4}\; ,
\end{eqnarray}
where the constants $f^\prime$ of $\mathfrak{G}_c$ are given in eq.~\eqref{3str}.

We can see in eqs.~\eqref{gc-1}-\eqref{gc-4} that the elements in $\mathcal{W}^{\prime(0)}=
\langle ad_{\mathcal{X}^{\prime (0)}}\rangle \subset \mathrm{Lie}\,\mathfrak{G}_c$ determine
a subalgebra, denoted $\mathcal{W}^{\prime(0)}$ as its vector
space. $\mathcal{W}^{\prime(0)}$ is therefore the Lie algebra associated with the Filippov subalgebra
$\mathfrak{G}_0 \subset \mathfrak{G}_c$, and acts on the coset
${\mathcal{W}^{\prime (1)}} = \textrm{Lie}\,\mathfrak{G}_c / \mathcal{W}^{\prime(0)}$ which is an abelian
ideal of Lie$\,\mathfrak{G}_c$. Thus, Lie$\,\mathfrak{G}_c$ has the semidirect structure
Lie$\,\mathfrak{G}_c={\mathcal{W}^{\prime (1)}} \, {\supset \!\!\!\!\!\! \raisebox{1.5pt} {\tiny +}}
\;\,{\mathcal{W}^{\prime (0)}}$ and dim$\, \mathrm{Lie}\,\mathfrak{G}_c=\left(
                                                             \begin{array}{c}
                                                             \mathrm{dim}\,\mathfrak{G}_c \\
                                                              2 \\
                                                             \end{array}
                                                                        \right)
-\mathrm{dim}\,(\mathrm{ker}\,ad)$,
where $ad: \wedge^2 \mathfrak{G}_c \rightarrow \mathrm{End}\, \mathfrak{G}_c$.

\subsubsection{Example: the contractions of $A_4$ and their associated Lie$\,(A_4)_c$}
\label{sec.A4c}

The simple euclidean FA $A_4$ (eq.~\eqref{A4}) has two possible types of non-trivial
subalgebras $ \mathfrak{G}_0$: one-dimensional, generated by any one
element of $A_4$, and two-dimensional, generated by any two elements
of the basis of $A_4$. They are both abelian, $[\mathfrak{G}_0,
\mathfrak{G}_0, \mathfrak{G}_0] =0$.

\begin{itemize}
\item{ \bf First case: $\mathfrak{G}_0$ one-dimensional}

Let $\mathfrak{G}_0$ be generated by $X_{4}$; the basis of
$\mathfrak{V}$ is then $\{X_{1},X_{2},X_{3}\}$ and
$\mathfrak{G}=\mathfrak{G}_0 \oplus \mathfrak{V}$.\\

{\bf a) Contraction}

If $\mathfrak{G}_0=\langle X_{4}\rangle$, eqs.~\eqref{3liegcstr-1}, \eqref{3liegcstr-2} show that
the contraction of $A_4$ with respect to $\mathfrak{G}_0$
gives rise to a four-dimensional abelian FA $(A_4)_c$. $\mathfrak{V}$ is then a subalgebra
of $\mathfrak{G}_c$ acting trivially on $\mathfrak{G}_0$. \\

{\bf b) Lie$\,(A_4)_c$}

Since $(A_4)_c$ is abelian, all $f^\prime_{j_1 \, j_2 \, j_3}{}^{j_4} = 0$,
$ad_{\mathcal{X}^\prime} \in \mathrm{ker}\,ad$, $\forall \mathcal{X}^\prime \in \wedge^2(A_4)_c$
and Lie$\, (A_4)_c$ reduces to the zero derivation. \\

\item{ \bf Second case: $\mathfrak{G}_0$ bidimensional}

Let $\mathfrak{G}_0$ be now generated  by two elements, $\{X_a, \; a=1,2 \}$ say,
of the basis of $\mathfrak{G}$. Thus, in
$\mathfrak{G}=\mathfrak{G}_0 \oplus \mathfrak{V}$, the vector space
$\mathfrak{V}$ is generated by $\{X_u, \; u=3,4 \}$. Clearly, $\mathfrak{G}_0$ and
$\mathfrak{V}$ play in this case a similar
role, and eq.~\eqref{A4} gives
\begin{eqnarray}
&&[\mathfrak{G}_0,\mathfrak{G}_0,\mathfrak{G}_0]=0 \\
&&[\mathfrak{G}_0,\mathfrak{G}_0,\mathfrak{V}] \subset \mathfrak{V} \\
&&[\mathfrak{G}_0,\mathfrak{V},\mathfrak{V}]\subset \mathfrak{G}_0 \\
&&[\mathfrak{V},\mathfrak{V},\mathfrak{V}]=0 \, .
\end{eqnarray}
Thus, $\mathfrak{G}_0$ and $\mathfrak{V}$ play a symmetrical role,
and both determine two-dimensional abelian Filippov subalgebras.

{\bf a) Contraction}

The only structure constants of $(A_4)_c$ different
from zero are, from eq.~\eqref{3str},
\begin{eqnarray}
f^\prime_{a_1a_2u_1}{}^{u_2}=\epsilon_{a_1a_2u_1}{}^{u_2} \,.
\end{eqnarray}
Therefore all the commutators in $(A_4)_c$ are zero except those coming from
$[\mathfrak{G}_0,\mathfrak{G}_0,\mathfrak{V}] \subset \mathfrak{V}$,
\begin{eqnarray}\label{A4c-1}
&&[X^\prime_{a_1},X^\prime_{a_2},X^\prime_{a_3}]=0 \\
&&[X^\prime_{a_1},X^\prime_{a_2},X^\prime_{u_1}]=\epsilon_{a_1a_2u_1}{}^{u_2}
X^\prime_{u_2} \label{A4c-2}\\
&&[X^\prime_{a_1},X^\prime_{u_1},X^\prime_{u_2}]=0 \label{A4c-3}\\
&&[X^\prime_{u_1},X^\prime_{u_2},X^\prime_{u_3}]=0  \label{A4c-4}\, ,
\end{eqnarray}
{\it i.e.}, except
\begin{equation}
[X^\prime_{1},X^\prime_{2},X^\prime_{3}]=X^\prime_4 \quad ,
\quad [X^\prime_{1},X^\prime_{2},X^\prime_{4}]=- X^\prime_3 \, .
\end{equation}

 The inner derivation associated with
$\mathcal{X}^\prime \in \wedge^2 \mathfrak{G}_0$,
$\mathfrak{G}_0\subset (A_4)_c$ acts on the two-dimensional abelian
ideal $\mathfrak{V}\subset (A_4)_c$ as
a $so(2)$ rotation.\\

{\bf b) Lie$\,(A_4)_c$}

To find the associated Lie~$(A_4)_c$, with $(A_4)_c$ given by eqs.~\eqref{A4c-1}-\eqref{A4c-4},
let us consider the vector spaces generated by the $ad_{\mathcal{X}^\prime}$ when the
$\mathcal{X}^\prime \in \wedge^2 (A_4)_c$
are labelled according to the pattern above. This leads to
\begin{eqnarray}\label{spLieA4c}
&& {\mathcal{W}^{\prime(0)}}= \langle ad_{(X^\prime_{a_1}, X^\prime_{a_2})}\rangle
= \langle ad_{(X^\prime_{1}, X^\prime_{2})}\rangle \, , \; \nonumber \\
&& {\mathcal{W}^{\prime(1)}}= \langle ad_{(X^\prime_{a}, X^\prime_{u})}\rangle  =
\langle ad_{(X^\prime_{1}, X^\prime_{3})},ad_{(X^\prime_{1}, X^\prime_{4})},
ad_{(X^\prime_{2}, X^\prime_{3})},ad_{(X^\prime_{2}, X^\prime_{4})}\rangle  \, , \; \nonumber \\
&& {\mathcal{W}^{\prime(2)}}= \langle ad_{(X^\prime_{u_1}, X^\prime_{u_2})}\rangle  =
\langle ad_{(X^\prime_{3}, X^\prime_{4})}\rangle  = \{0\} \, .
\end{eqnarray}
Then, applying eqs.~\eqref{gc-1}-\eqref{gc-4} to this case, we find that
Lie$\,(A_4)_c$ is given by the commutators
\begin{eqnarray} \label{LieA4c-1}
&& [ad_{\mathcal{X}_{a_1 a_2}^{\prime (0)}}, ad_{\mathcal{Y}_{b_1 b_2}^{\prime (0)}}]  =  0 \\
&& [ad_{\mathcal{X}_{a_1a_2}^{\prime (0)}}, ad_{\mathcal{Y}_{b_1u_2}^{\prime (1)}}]
= \frac12 \epsilon_{{a_1}\, {a_2}\, {u_2}}{}^{u} ad_{(X^\prime_{b_1}, X^\prime_u)}-
\frac12 \epsilon_{{b_1}\, {u_2}\, {a_1}}{}^{u} ad_{(X^\prime_{u}, X^\prime_{a_2})} \nonumber\\
&& \qquad - \frac12 \epsilon_{{b_1}\, {u_2}\, {a_2}}{}^{u} ad_{(X^\prime_{a_1}, X^\prime_u)} \quad
  \in \mathcal{W}^{\prime(1)} \label{LieA4c-2}\\
&& [ad_{\mathcal{X}_{a_1u_1}^{\prime (1)}}, ad_{\mathcal{Y}_{a_2 u_2}^{\prime (1)}}]
= \frac12 \epsilon_{{a_1}\, {u_1}\, {a_2}}{}^{u} \underbrace{ad_{(X^\prime_{u}, X^\prime_{u_2})}}_{=0}
- \frac12 \epsilon_{{a_2}\, {u_2}\, {a_1}}{}^{u} \underbrace{ad_{(X^\prime_{u},
X^\prime_{u_1})}}_{=0} =0  \label{LieA4c-3} \\
&& [ad_{\mathcal{X}_1^{\prime (2)}}, ad_{\mathcal{Y}_2^{\prime (r)}}]=0\;,
\quad r=0,1,2 \; . \label{LieA4c-4}
\end{eqnarray}
The $r.h.s.$ $\epsilon_{{a_1}\, {a_2}\, {u_2}}{}^{u}
ad_{(X^\prime_{b_1}, X^\prime_u)}$ of eq.~\eqref{LieA4c-2} is
non-zero when $b_1=a_1$ or $b_1=a_2$, and the $r.h.s.$ of
eq.~\eqref{LieA4c-3} is always zero since $ad_{(X^\prime_{u},
X^\prime_{v})}=0$. As shown in Sec.~\ref{sec.3LieGc},
${\mathcal{W}^{\prime(0)}} \subset \textrm{Lie}\,\mathfrak{G}_c$ is
a subalgebra, abelian in this case, that acts on the abelian ideal
$\mathcal{W}^{\prime(1)}\subset \textrm{Lie}\,\mathfrak{G}_c$. Thus,
Lie$\,(A_4)_c$ has the semidirect structure
Lie$\,(A_4)_c={{\mathcal{W}^{\prime (1)}} \, {\supset \!\!\!\!\!\!
\raisebox{1.5pt} {\tiny +}} \;\,{\mathcal{W}^{\prime (0)}}} $, and
is the five-dimensional Lie algebra $(Tr_2\oplus Tr_2) \, {\supset
\!\!\!\!\!\! \raisebox{1.5pt} {\tiny +}} \;\, so(2)$ where $so(2)$
acts independently on the two bidimensional abelian subalgebras
$\langle ad_{\mathcal{X}_{13}^{\prime
(1)}},ad_{\mathcal{X}_{14}^{\prime (1)}}\rangle$, $\langle
ad_{\mathcal{X}_{23}^{\prime (1)}},ad_{\mathcal{X}_{24}^{\prime
(1)}}\rangle$ (translations $Tr_2$) of $\mathcal{W}^{\prime(1)}$. We
check that dim$\,\mathrm{Lie} \,(A_4)_c=6-1=5$ since
 $\mathcal{X}_{34}^{\prime (2)} \in \mathrm{ker}\, ad$.  \\

\end{itemize}

\subsection{General case: contractions of $n$-Lie algebras $\mathfrak{G}$}
\label{Sec.Gc}
Having discussed the $n=3$ case it is not difficult to
extend the contraction procedure to an arbitrary $n$-Lie algebra
$\mathfrak{G}$ (eq.~\eqref{nFIdef}). Let $\mathfrak{G}_0$ now be a
subspace of $\mathfrak{G}$ (not yet a subalgebra) and split the
vector space of $\mathfrak{G}$ as the sum
\begin{equation}\label{splittingnLieI}
\mathfrak{G}=\mathfrak{G}_0 \oplus \mathfrak{V}\; ,\quad
\begin{array}{l}
\{X_{a}\}\;\mathrm{basis\, of \,} \mathfrak{G}_0 \, ,  \; a \in I_0 =
\{1,\ldots, \textrm{dim} \mathfrak{G}_0\} \quad \,\\
\{X_{u}\} \; \mathrm{basis\, of \,} \mathfrak{V} \,,  \; u \in I_1 =
\{\textrm{dim} \mathfrak{G}_0 + 1,\ldots, \textrm{dim} \mathfrak{G} \} \quad \\
\end{array}
\end{equation}
where, again, the indices $a$, $b$, $c$ refer here to the basis of $\mathfrak{G}_0$,
$u$, $v$, $w$ to the basis of $\mathfrak{V}$ and $j$, $k$, $l$ label the elements
of the basis of the FA $\mathfrak{G}$,
$$ X_a, X_b, X_c \in \mathfrak{G}_0 \subset \mathfrak{G} \,,
\quad X_u, X_v, X_w \in \mathfrak{V} \subset \mathfrak{G} \,,\quad X_j, X_k, X_l \in \mathfrak{G} \,. $$
Then, an arbitrary $n$-Lie bracket in $\mathfrak{G}$ may be written as
\begin{eqnarray}
[X_{a_1}, \ldots, X_{a_p} , X_{u_{p+1}}, \ldots X_{u_n}]
&=& f_{a_1 \ldots a_p \,u_{p+1} \ldots u_n}{}^l X_l = \nonumber \\
&=& f_{a_1 \ldots a_p \,u_{p+1} \ldots u_n}{}^a X_a
+ f_{a_1 \ldots a_p \,u_{p+1} \ldots u_n}{}^u X_u \, .\quad
\end{eqnarray}

Let us rescale the basis generators of $\mathfrak{V}$, $X_u \rightarrow X_u^\prime
\equiv \epsilon X_u$ while keeping those of $\mathfrak{G}_0$ unscaled,
$X_a \rightarrow X_a^\prime = X_a$. Then,
\begin{eqnarray}
&&[X^\prime_{a_1}, \ldots, X^\prime_{a_p} , X^\prime_{u_{p+1}}, \ldots ,  X^\prime_{u_n}]=
\nonumber \\
&& = \epsilon^{n-p} (f_{a_1 \ldots a_p \,u_{p+1} \ldots u_n}{}^a X_a
+ f_{a_1 \ldots a_p \, u_{p+1} \ldots u_n}{}^u X_u) = \nonumber \\
&& = \epsilon^{n-p} f_{a_1 \ldots a_p \,u_{p+1} \ldots u_n}{}^a X^\prime_a
+ \epsilon^{n-p-1} f_{a_1 \ldots a_p \, u_{p+1} \ldots u_n}{}^u  X^\prime_u.
\label{limit}
\end{eqnarray}
\noindent The limit $\epsilon \rightarrow 0$ is well defined for the
first term in the last equality because $n \geq p$ always, but to
have a well defined limit for the second one when $n=p$ we must have
$f_{a_1 \ldots a_n}{}^u=0$ so that the factor $\epsilon^{-1}$ does
not appear. Therefore, $\mathfrak{G}_0$ must be a subalgebra of
$\mathfrak{G}$: {\it FA contractions $\mathfrak{G}_c$ have to be
defined with respect to Filippov subalgebras $\mathfrak{G}_0 \subset
\mathfrak{G}$}.

The limit $\epsilon \rightarrow 0$ defines the contraction $\mathfrak{G}_c$
of the $n$-Lie algebra $\mathfrak{G}$ with respect to its subalgebra $\mathfrak{G}_0$.
The $n$-brackets of $\mathfrak{G}_c$  are given by:
\begin{equation}
{}[X^\prime_{a_1}, \ldots, X^\prime_{a_p} , X^\prime_{u_{p+1}}, \ldots  X^\prime_{u_n}]
= f^{\prime}_{a_1 \ldots a_p \,u_{p+1} \ldots u_n}{}^l X^\prime_l\, ,
\end{equation}
where
\begin{equation}
f^{\prime}_{a_1\, \ldots\, a_p \,u_{p+1} \ldots\, u_n}{}^l=
\left\{
\begin{array}{l}
\textrm{lim}_{\epsilon\rightarrow 0}\; \epsilon^{n-p}
f_{a_1\, \ldots\, a_p \,u_{p+1} \ldots\, u_n}{}^a \;,
\quad \;\; \forall a\in I_0 \\
\textrm{lim}_{\epsilon\rightarrow 0}\; \epsilon^{n-p-1}
f_{a_1\, \ldots\, a_p \,u_{p+1} \ldots \, u_n}{}^u \;,
\quad \forall u \in I_1 \quad .\\
\end{array}
\right.
\end{equation}

\noindent Therefore, the structure constants of $\mathfrak{G}_c$ are given by

\begin{equation}
\label{str}
f^{\prime}_{a_1, \ldots, a_p ,u_{p+1} \ldots, u_n}{}^l=
\left\{ \!\!
\begin{array}{ll} \!
\left.
\begin{array}{lll}
f^\prime_{a_1 \ldots a_n}{}^a & \quad \qquad \;\;=
\!  f_{a_1 \ldots a_n}{}^a \;, & \qquad  p=n,\;\; a\in I_0,   \\
f^\prime_{a_1 \ldots a_n}{}^u & \quad \qquad \;\; = \! 0 \;, & \qquad
p=n,\;\; u\in I_1,   \\
\end{array}
\right\} & \;\! \! \! \! (a) \\ \!
\left.
\begin{array}{lll}
f^\prime_{a_1\, \ldots \, a_{n-1} \,u_n}{}^a &\!
= 0 &  p=n-1,\;\; a \in I_0\\
f^\prime_{a_1\, \ldots \, a_{n-1} \,u_n}{}^u &\!
= f_{a_1\, \ldots \, a_{n-1} \,u_n}{}^u \;, & p=n-1,\;\; u \in I_1\\
f^{\prime}_{a_1, \ldots, a_p ,u_{p+1} \ldots, u_n}{}^l &\!
= 0 &  p<n-1,\;\; l \in I_0 \cup I_1 \\
\end{array}
\right. & \!\!\! \! \begin{array}{l}(b) \\
(c) \\ (d) \\ \end{array} \\
\end{array}
\right.
\end{equation}
Again (eq.~(\ref{str}a)), the Filippov subalgebra $\mathfrak{G}_0$
is preserved in the contraction. For $n=3$, eqs.~\eqref{str}
reproduce eq.~\eqref{3str}.

To see that the structure constants of (\ref{str}) define indeed an
$n$-Lie algebra $\mathfrak{G}_c$, we have to check the Filippov
identity for $\mathfrak{G}_c$. As expected, this is satisfied as a
consequence of the FI for the original FA $\mathfrak{G}$. Indeed,
the FI for the contracted algebra,
\begin{eqnarray}
\label{FI}
[X^\prime_1,\ldots, X^\prime_{n-1}, [Y^\prime_1,\ldots, Y^\prime_n]]
=\sum_{i=1}^n [Y^\prime_1\dots Y^\prime_{i-1},
[X^\prime_1, \ldots, X^\prime_{n-1},Y^\prime_i],Y^\prime_{i+1},\ldots, Y^\prime_n]
\end{eqnarray}
gives, in term of the primed structure constants of $\mathfrak{G}_c$
\begin{equation}
\label{FIstr}
f^\prime_{k_1\ldots k_n}{}^{i} f^\prime_{l_1\dots l_{n-1}i}{}^j
=\sum_{i=1}^n f^\prime_{l_1\ldots l_{n-1} k_i}{}^{i}
f^\prime_{k_1\ldots k_{i-1}i k_{i+1}\ldots k_n}{}^j \quad .
\end{equation}
The proof involves three possible cases:
\begin{enumerate}{\leftmargin=-1em}
\item
All algebra elements in (\ref{FI}) belong to $\mathfrak{G}_0$. \\
\noindent Then the structure constants in (\ref{FIstr}) are given by
eq.~(\ref{str}a), and the FI holds because $\mathfrak{G}_0 \subset
\mathfrak{G}_c$ is a Filippov (sub)algebra.
\item
Only one element in (\ref{FI}) belongs to $\mathfrak{V}$,
and the remaining $2n-2$ ones belong to $\mathfrak{G}_0$. \\
 \noindent
In this case, when the index $j\in I_0$ in \eqref{FIstr},
we have the identity $0=0$. Indeed, due to (\ref{str}a),
(\ref{str}b), all the indices in the terms $f^\prime_{--}{}^j$ must
belong to $I_0$ to be non-zero, but then the structure constants
$f^\prime_{--}{}^i$ are of the type (\ref{str}b),
and therefore vanish. \\
When $j\in I_1$, the FI is the same for the contracted
$\mathfrak{G}_c$ and the original $n$-Lie algebra $\mathfrak{G}$.
The reason is that in this case the terms that may be non-zero in
the FI \eqref{FIstr} are the same for $\mathfrak{G}_c$ and
$\mathfrak{G}$ and involve structure constants of the type
$f^\prime_{a_1\ldots a_{n-1}u}{}^v=f_{a_1 \ldots a_{n-1}u}{}^v$
since $f^\prime_{a_1 \ldots a_n}{}^u=0$ for both $\mathfrak{G}$ and
$\mathfrak{G}_c$.
\item
Two or more elements belong to $\mathfrak{V}$. \\
\noindent In this case, as in the previous one, when $j\in I_0$, we
have the identity $0=0$ because due to (\ref{str}a), (\ref{str}b),
(\ref{str}d) all the indices in the structure constants
$f^\prime_{--}{}^j$ must be in $I_0$ to be non-zero, but then the
other structure constants in the products are of the form
(\ref{str}d), and therefore are zero. When $j\in I_1$ we have again
$0=0$, because in this case $f^\prime_{--}{}^j$ has to be of the
form (\ref{str}c) to be non-zero, and then the terms
 $f^\prime_{--}{}^i$ are either of the form (\ref{str}b) if $i\in I_0$
 or (\ref{str}d) if $i\in I_1$, which vanish in both cases.
\end{enumerate}

The $n$-brackets of the contraction $\mathfrak{G}_c$ of the FA
$\mathfrak{G}$ with respect to the subalgebra $\mathfrak{G}_0$
 have therefore the following general structure (see eq.~\eqref{str}):
\begin{equation}\label{str1}
\begin{array}{lll}
 [\mathfrak{G}_0, \ldots, \mathfrak{G}_0] & \subset \mathfrak{G}_0,
 & \quad (ad_{\mathcal{X}_0} \mathfrak{G}_0  \subset \mathfrak{G}_0) \\
{} [\mathfrak{G}_0, \ldots, \mathfrak{G}_0, \mathfrak{V}] & \subset \mathfrak{V},
&  \quad ( ad_{\mathcal{X}_0} \mathfrak{V} \subset \mathfrak{V} ) \\
{} [\mathfrak{G}_0, \ldots, \mathfrak{G}_0, \mathfrak{V}, \mathfrak{V}] & =0, &  \\
{} \ldots \ldots && \\
{} [\mathfrak{G}_0, \mathfrak{V}, \ldots, \mathfrak{V}] & =0, &  \\
{} [\mathfrak{V}, \ldots,  \mathfrak{V}] & =0, &  \\
\end{array}
\end{equation}
where $\mathcal{X}_0=(X_1,\ldots,X_{n-1})$, $X_1,\ldots,X_{n-1}\in \mathfrak{G}_0$.
The elements in the coset $\mathfrak{V}=\mathfrak{G}/\mathfrak{G}_0$ become an abelian ideal in
$\mathfrak{G}_c$, $[\mathfrak{G}, \ldots, \mathfrak{G}, \mathfrak{V}] \subset \mathfrak{V}$,
$[\mathfrak{V}, \ldots, \mathfrak{V}] =0 $, and the fundamental objects of
$\mathfrak{G}_0 \subset \mathfrak{G}_c$ act on $\mathfrak{V}$ as derivations,
$ad_{\mathcal{X}_0}: \mathfrak{V} \rightarrow \mathfrak{V}$. Thus, $\mathfrak{G}_c$
{\it has the FA semidirect structure} $\mathfrak{G}_c={\mathfrak{V}}\,
{\supset \!\!\!\!\!\! \raisebox{1.5pt} {\tiny +}} \;\,{\mathfrak{G}_0}$.

\subsubsection{The Lie algebra Lie$\,\mathfrak{G}_c$ associated with a contraction $\mathfrak{G}_c$}\label{sec.LieGc}

To describe Lie$\,\mathfrak{G}_c$ associated with $\mathfrak{G}_c$,
it will prove again useful to split the space of fundamental objects
of $\mathfrak{G}_c$ in  subsets, where each subset $\mathcal{X}^{\prime (r)}$
is characterized by the number $r$ of elements $X^\prime_{u_i}\in \mathfrak{V}$
in the $\mathcal{X}^{\prime}$s that it contains. Thus,
\begin{eqnarray}
&&\mathcal{X}^{\prime (0)}= (X^\prime_{a_1}, \ldots, X^\prime_{a_{n-1}}) \, , \quad
X^\prime_{a_i} \in \mathfrak{G}_0 \label{subsetliegc1}\\
&&\mathcal{X}^{\prime (1)}= (X^\prime_{a_1}, \ldots, X^\prime_{a_{n-2}}, X^\prime_{u_{n-1}}) \, , \quad
X^\prime_{a_i} \in \mathfrak{G}_0 \, , \; X^\prime_{u_i}\in \mathfrak{V}\\
&&\cdots \nonumber\\
&&\mathcal{X}^{\prime (r)}= (X^\prime_{a_1}, \ldots, X^\prime_{a_{n-r-1}},
X^\prime_{u_{n-r}}, \ldots, X^\prime_{u_{n-1}})
 \, , \quad X^\prime_{a_i} \in \mathfrak{G}_0 \, , \; X^\prime_{u_i}\in \mathfrak{V}\\
&&\cdots \nonumber\\
&&\mathcal{X}^{\prime (n-1)}= (X^\prime_{u_1}, \ldots, X^\prime_{u_{n-1}}) \, ,
\quad X^\prime_{u_i}\in \mathfrak{V} \quad, \label{subsetliegc5}
\end{eqnarray}
and the vector spaces generated by the inner derivations associated with the
fundamental objects in $\mathcal{X}^{\prime(r)}$ are denoted by
\begin{equation}\label{LieGvspaces}
\mathcal{W}^{\prime(r)}=\langle ad_{\mathcal{X}^{\prime(r)}}\rangle\;, \quad r=0,\ldots, n-1 \; .
\end{equation}
Due to eqs.~\eqref{str}, we see that the inner derivations of Lie$\,\mathfrak{G}_c$
act on the elements of the contracted $\mathfrak{G}_c$ in the following way:
\begin{eqnarray} \label{str-ad}
&&ad_{\mathcal{X}^{\prime (0)}} \mathfrak{G}_0 \subset \mathfrak{G}_0 \\
&&ad_{\mathcal{X}^{\prime (0)}} \mathfrak{V}\subset \mathfrak{V} \label{str-ad2}\\
&&ad_{\mathcal{X}^{\prime (1)}} \mathfrak{G}_0\subset \mathfrak{V} \label{str-ad3}\\
&&ad_{\mathcal{X}^{\prime (1)}} \mathfrak{V} = 0 \label{str-ad4}\\
&&ad_{\mathcal{X}^{\prime (r)}} \mathfrak{G}_c = 0 \quad \forall r\ge 2 \, \label{str-ad5}
\end{eqnarray}
(eqs.~\eqref{str-ad2}, \eqref{str-ad3} both correspond to the second equation in \eqref{str1}).
Therefore $ad_{\mathcal{X}^{\prime (r)}} = 0$ for $r\ge 2$ {\it i.e.}, when $r\ge 2$ all the
$\mathcal{X}^{\prime(r)}$ belong to ker$\,ad$ and $\mathcal{W}^{\prime(r)}=0$.

The composition of fundamental objects in eq.~\eqref{comp} and the structure constants \eqref{str} of
the contracted FA $\mathfrak{G}_c$  determine the following structure for Lie$\,\mathfrak{G}_c$
\begin{eqnarray}
&& [ad_{\mathcal{X}_{{a_1}\,\ldots\, {a_{n-1}}}^{\prime (0)}},
ad_{\mathcal{Y}_{{b_1}\,\ldots\, {b_{n-1}}}^{\prime (0)}}]
= \frac12 ad_{[(X^\prime_{a_1}, \ldots, X^\prime_{a_{n-1}}) \cdot
(X^\prime_{b_1}, \ldots, X^\prime_{b_{n-1}})
- (X^\prime_{b_1}, \ldots, X^\prime_{b_{n-1}}) \cdot
(X^\prime_{a_1}, \ldots, X^\prime_{a_{n-1}})]}= \nonumber\\
&& \quad = \frac12 ad_{\left[\sum_{i=1}^{n-1} (X^\prime_{b_1}, \ldots,
[X^\prime_{a_1},\ldots, X^\prime_{a_{n-1}}, X^\prime_{b_{i}}] , \ldots, X^\prime_{b_{n-1}})
- (a\leftrightarrow b) \right]} = \qquad \nonumber\\
&& \quad \frac12 \left( \sum_{i=1}^{n-1} f_{{a_1}\,\ldots\, {a_{n-1}}\, {b_i}}{}^{b}
ad_{(X^\prime_{b_1}, \ldots, {X}^\prime_{b_{i-1}},
X^\prime_b, {X}^\prime_{b_{i+1}} \ldots, X^\prime_{b_{n-1}})}
- (a\leftrightarrow b)\right) \quad \in {\mathcal{W}^{\prime (0)}} \label{nLieG00}
\\[0.5cm]
&& [ad_{\mathcal{X}_{{a_1}\,\ldots\, {a_{n-1}}}^{\prime (0)}},
ad_{\mathcal{Y}_{{b_1}\,\ldots\, {b_{n-2}} \, {v_{n-1}}}^{\prime (1)}}]=  \nonumber\\
&& \quad = \frac12 ad_{[(X^\prime_{a_1}, \ldots, X^\prime_{a_{n-1}})
\cdot (X^\prime_{b_1}, \ldots, X^\prime_{b_{n-2}}, X^\prime_{v_{n-1}})
- (X^\prime_{b_1}, \ldots, X^\prime_{b_{n-2}}, X^\prime_{v_{n-1}})
\cdot (X^\prime_{a_1}, \ldots, X^\prime_{a_{n-1}})]} = \quad \nonumber\\
&& \quad \frac12 \left(\sum_{i=1}^{n-2} f_{{a_1}\,\ldots\, {a_{n-1}}\, {b_i}}{}^{b}
 ad_{(X^\prime_{b_1}, \ldots, {X}^\prime_{b_{i-1}}, X^\prime_b, {X}^\prime_{b_{i+1}},
 \ldots, X^\prime_{b_{n-2}}, X^\prime_{v_{n-1}})}
  + f_{{a_1}\,\ldots\, {a_{n-1}}\, {v_{n-1}}}{}^{v}
ad_{(X^\prime_{b_1}, \ldots, X^\prime_{b_{n-2}}, X^\prime_v)} \right.  \quad \nonumber \\
&& \left. \quad - \sum_{i=1}^{n-1} f_{{b_1}\,\ldots\, {b_{n-2}} \, {v_{n-1}} \, {a_i}}{}^{v}
ad_{(X^\prime_{a_1}, \ldots, {X}^\prime_{a_{i-1}}, X^\prime_v,
{X}^\prime_{a_{i+1}}, \ldots, X^\prime_{a_{n-1}})} \right)
\quad \in {\mathcal{W}^{\prime (1)}} \label{nLieG01}
\\[0.5cm]
&& [ad_{\mathcal{X}_{{a_1}\,\ldots\, {a_{n-1}}}^{\prime (0)}},
ad_{\mathcal{Y}_{{b_1}\,\ldots\, {b_{n-3}} \, {v_{n-2}} \, {v_{n-1}}}^{\prime (2)}}] = \nonumber\\
&& \quad = \frac12 ad_{[(X^\prime_{a_1}, \ldots, X^\prime_{a_{n-1}})
\cdot (X^\prime_{b_1}, \ldots, X^\prime_{b_{n-3}}, X^\prime_{v_{n-2}},
X^\prime_{v_{n-1}})- (X^\prime_{b_1}, \ldots, X^\prime_{b_{n-3}},
X^\prime_{v_{n-2}}, X^\prime_{v_{n-1}})
\cdot (X^\prime_{a_1}, \ldots, X^\prime_{a_{n-1}})]} \quad \nonumber\\
&& \quad =\frac12 \left(\sum_{i=1}^{n-3} f_{{a_1}\,\ldots\, {a_{n-1}}\, {b_i}}{}^{b}
ad_{(X^\prime_{b_1}, \ldots, {X}^\prime_{b_{i-1}}, X^\prime_b, {X}^\prime_{b_{i+1}}, \ldots,
X^\prime_{b_{n-3}}, X^\prime_{v_{n-2}}, X^\prime_{v_{n-1}})}
\right.\quad \nonumber\\
&& \quad \left. + f_{{a_1}\,\ldots\, {a_{n-1}}\, {v_{n-2}}}{}^{v}
ad_{(X^\prime_{b_1}, \ldots, X^\prime_{b_{n-3}}, X^\prime_{v_{n-1}}, X^\prime_v)}
+ f_{{a_1}\,\ldots\, {a_{n-1}}\, {v_{n-1}}}{}^{v}
ad_{(X^\prime_{b_1}, \ldots, X^\prime_{b_{n-3}}, X^\prime_v, X^\prime_{v_{n-2}})}\right)  \; =0  \qquad
\\[0.5cm]
&& [ad_{\mathcal{X}_{{a_1}\,\ldots\, {a_{n-2}} \, {u_{n-1}}}^{\prime (1)}},
ad_{\mathcal{Y}_{{b_1}\,\ldots\, {b_{n-2}} \, {v_{n-1}}}^{\prime (1)}}] = \nonumber\\
&& \quad = \frac12 ad_{\left[(X^\prime_{a_1}, \ldots, X^\prime_{a_{n-2}},X^\prime_{u_{n-1}})
\cdot (X^\prime_{b_1}, \ldots, X^\prime_{b_{n-2}}, X^\prime_{v_{n-1}})
- (X^\prime_{b_1}, \ldots, X^\prime_{b_{n-2}},X^\prime_{v_{n-1}})
\cdot (X^\prime_{a_1}, \ldots, X^\prime_{a_{n-2}}, X^\prime_{u_{n-1}}) \right]} \nonumber\\
&& \quad = \frac12 \left(\sum_{i=1}^{n-2} f_{{a_1}\,\ldots\, {a_{n-2}}\,{u_{n-1}}\, {b_i}}{}^{v}
ad_{(X^\prime_{b_1}, \ldots, {X}^\prime_{b_{i-1}}, X^\prime_v, {X}^\prime_{b_{i+1}},
\ldots, X^\prime_{b_{n-2}}, X^\prime_{v_{n-1}})}
- (a,u \leftrightarrow b,v) \right) \; =0 \;
\\[0.5cm]
&& [ad_{\mathcal{X}^{\prime (r)}}, ad_{\mathcal{Y}^{\prime (2)}}]= 0\, , \quad r = 1, 2 \label{nLieG0n}
\end{eqnarray}
where we have used eqs.~\eqref{str-ad4}, \eqref{str-ad5} and the only non-zero structure constants
appearing above are the $f^\prime_{{a_1}\,\ldots\, {a_{n}}}{}^{a}=f_{{a_1}\,\ldots\, {a_{n}}}{}^{a}$
and $f^\prime_{{a_1}\,\ldots\, {a_{n-1}}\, {u_{n}}}{}^{u}=f_{{a_1}\,\ldots\, {a_{n-1}}\, {u_{n}}}{}^{u}$
by eq.~\eqref{str}. As a result, the structure of
Lie~$\mathfrak{G}_c$ for the $n$-Lie algebra $\mathfrak{G}_c$ is similar to the one
found for $n=3$. The elements $ad_{\mathcal{X}^{\prime (0)}}$ generate
a subalgebra $\mathcal{W}^{\prime(0)}$ of Lie$\,\mathfrak{G}_c$ and the
$ad_{\mathcal{X}^{\prime (1)}}$ an abelian ideal $\mathcal{W}^{\prime(1)}$.
Lie$\,\mathfrak{G}_c$ has therefore the semidirect structure
$\mathrm{Lie}\,\mathfrak{G}_c = \mathcal{W}^{\prime(1)} \,  {\supset \!\!\!\!\!\! \raisebox{1.5pt} {\tiny +}}
\;\,\mathcal{W}^{\prime(0)}$ which, for $n=3$, recovers the case of Sec.~\ref{sec.3LieGc}.
As for any Lie algebra associated with a FA,
dim$\,\mathrm{Lie} \,\mathfrak{G}_c=\left(
                                       \begin{array}{c}
                                              \mathrm{dim}\,\mathfrak{G}_c \\
                                              n-1 \\
                                       \end{array}
                                    \right)
                                   -\mathrm{dim}\,(\mathrm{ker}\,ad)$
where now $ad:\wedge^{n-1}\mathfrak{G}_c \rightarrow \mathrm{End}\,\mathfrak{G}_c$.

\subsubsection{Example: the contractions of $A_{n+1}$ and their associated Lie$\,(A_{n+1})_c$ }
\label{Sec.An+1c}
In this section we consider the general simple Euclidean
FAs $\mathfrak{G}:=A_{n+1}$,
\begin{equation}
[X_{l_1},\ldots,X_{l_n}]=\epsilon_{l_1 \ldots l_n}{}^{l_{n+1}} X_{l_{n+1}}\; ,
\end{equation}
generalizing the $n=3$ results of Sec.~\ref{sec.A4c}. There are various
subspaces that determine different non-trivial subalgebras
$\mathfrak{G}_0\subset A_{n+1}$ and corresponding vector space
splittings $\mathfrak{G}=\mathfrak{G}_0 \oplus \mathfrak{V}$: it
suffices  to take $\mathfrak{G}_0$ generated by $m$ basis elements
of $A_{n+1}$ with $m\le (n-1)$ (if dim$\,\mathfrak{G}_0 =n$,
$\mathfrak{G}_0$ cannot be a subalgebra when $\mathfrak{G}$ is
simple). We shall see below that only one of these splittings, when
$m=n-1$, leads to a non-trivial contraction $(A_{n+1})_c$. All other
$(m< n-1)$-dimensional subalgebras lead to a contraction of
$A_{n+1}$ which is an abelian $(n+1)$-dimensional $n$-Lie algebra.

Let then $\mathfrak{G}_0$ be generated by $n-1$
basis elements of $A_{n+1}$ and $\mathfrak{V}$ by the remaining two,
\begin{equation}\label{Anspl}
\mathfrak{G}_0= \langle X_a, \; a=1,\ldots,n-1 \rangle, \quad
\mathfrak{V}=\langle X_u, \; u=n,n+1 \rangle .
\end{equation}
Then, the various $A_{n+1}$ $n$-brackets follow the pattern
\begin{eqnarray}
&&[\mathfrak{G}_0,\mathop{\dots}\limits^{n},\mathfrak{G}_0]=0 \\
&&[\mathfrak{G}_0,\mathop{\dots}\limits^{n-1},\mathfrak{G}_0,\mathfrak{V}]\subset \mathfrak{V} \\
&&[\mathfrak{G}_0,\mathop{\dots}\limits^{n-2},\mathfrak{G}_0,\mathfrak{V},\mathfrak{V}]
\subset \mathfrak{G}_0 \label{g..gvvg} \\
&&[\mathfrak{G}_0,\mathop{\dots}\limits^{n-3},\mathfrak{G}_0,\mathfrak{V},\mathfrak{V},\mathfrak{V}]=0 \\
&& \ldots \nonumber \\
&&[\mathfrak{V}, \mathop{\dots}\limits^{n},\mathfrak{V}]=0 \, .
\end{eqnarray}

Looking at eqs.~\eqref{str} we find that the only non-zero
structure constants of the contraction $(A_{n+1})_c$
of the FA $A_{n+1}$ with respect to a $(n-1)$-dimensional subalgebra are
\begin{eqnarray}
\label{An+1str}
f^\prime_{a_1\ldots a_{n-1} u_1}{}^{u_2}=\epsilon_{a_1\ldots a_{n-1} u_1}{}^{u_2} \; .
\end{eqnarray}
Note that any other ($m<n-1$)-dimensional $\mathfrak{G}_0$ would
lead to $f^\prime_{i_1\ldots i_{n}}{}^{k}=0 $. Thus, the splitting
\eqref{Anspl} is the only one leading to a non fully abelian
contraction.

The contracted $n$-Lie algebra $(A_{n+1})_c$ is given by
\begin{eqnarray}\label{An+1c-1}
&&[X^\prime_{a_1},\ldots,X^\prime_{a_n}]=0 \\
&&[X^\prime_{a_1},\ldots,X^\prime_{a_{n-1}},X^\prime_{u_1}]=
\epsilon_{a_1 \ldots a_{n-1} u_1}{}^{u_2} X^\prime_{u_2} \label{An+1c-2}\\
&&[X^\prime_{a_1},\ldots,X^\prime_{a_{n-2}},X^\prime_{u_1},X^\prime_{u_2}]=0 \label{An+1c-3}\\
&& \ldots \nonumber \\
&&[X^\prime_{u_1},\ldots,X^\prime_{u_n}]=0  \label{An+1c-4}\, .
\end{eqnarray}
It has a $(n-1)$-dimensional abelian subalgebra $\mathfrak{G}_0$
acting by eq.~\eqref{An+1c-2} on the two-dimensional abelian ideal
$\mathfrak{V}$. For $n=3$, this reproduces the contraction $(A_4)_c$
of the second case in Sec.~\ref{sec.A4c}.

The familiar Lie algebra case also follows in this framework. For
$n=2$, $A_3=so(3)$ and the ($n-1$)-dimensional subalgebra is of
dimension one. Then, the only non-zero structure constants in
eq.~\eqref{An+1str} reduce to $f^\prime_{au}{}^v=\epsilon_{au}{}^v$,
and $(A_3)_c= Tr_{2} \,{\supset \!\!\!\!\!\!\raisebox{1.5pt} {\tiny
+}} \;\, so(2)=E_2$,
the Euclidean algebra on the plane. \\

Let us now find Lie$\,(A_{n+1})_c$. For it, consider the adjoint maps
determined by the fundamental objects of $(A_{n+1})_c$ in the subsets
\eqref{subsetliegc1}-\eqref{subsetliegc5}, and the corresponding
vector spaces $\mathcal{W}^{\prime(r)}$ generated by them,
\begin{eqnarray}\label{spLieAn+1c}
&& {\mathcal{W}^{\prime(0)}}= \langle
ad_{(X^\prime_{a_1},\ldots, X^\prime_{a_{n-1}})}\rangle , \; \nonumber \\
&& {\mathcal{W}^{\prime(1)}}=
\langle ad_{(X^\prime_{a_1},\ldots, X^\prime_{a_{n-2}}, X^\prime_{u})}\rangle\, , \; \nonumber \\
&& {\mathcal{W}^{\prime(2)}}=
\langle ad_{(X^\prime_{a_1},\ldots, X^\prime_{a_{n-3}},X^\prime_{u_1}, X^\prime_{u_2})}\rangle
=\{0\}\, , \nonumber \\
&& \dots\dots  \nonumber \\
&& {\mathcal{W}^{\prime(n-1)}}=
\langle ad_{(X^\prime_{u_1},\ldots, X^\prime_{u_{n-1}})}\rangle =\{0\}  \quad ,
\end{eqnarray}
where, by eq.~\eqref{str-ad5}, $ad_{\mathcal{X}^{\prime (r)}}=0, \; r \ge 2$
so that $\mathcal{W}^{\prime(r)}=\{0\}$ for $r\ge 2$ (note that the non-zero
commutator in eq.~\eqref{g..gvvg} becomes zero in $(A_{n+1})_c$, eq.~\eqref{An+1c-3}).
Therefore, the vector space of Lie~$(A_{n+1})_c$ is reduced to
$\mathcal{W}^{\prime(0)}\oplus\mathcal{W}^{\prime(1)}$, of dimension
$\left(
  \begin{array}{c}
     n-1 \\
     n-1 \\
  \end{array}
\right)
+2\left(
\begin{array}{c}
 n-1 \\
 n-2 \\
\end{array}
\right) =2n-1$.

The structure of the Lie algebra Lie$\,(A_{n+1})_c$ is obtained by
inserting the structure constants $f^\prime_{l_1\ldots
l_n}{}^{l_{n+1}}$ of $(A_{n+1})_c$ (as given in eqs.~\eqref{str}
whith $f_{l_1\ldots l_n}{}^{l_{n+1}}=\epsilon_{l_1\ldots
l_n}{}^{l_{n+1}}$ since $\mathfrak{G}=A_{n+1}$) in
eqs.~\eqref{nLieG00}-\eqref{nLieG0n}. This leads to

\begin{eqnarray}\label{LieA4cn+1,0,0}
&& [ad_{\mathcal{X}_{{a_1}\,\ldots\, {a_{n-1}}}^{\prime (0)}},
ad_{\mathcal{Y}_{{b_1}\,\ldots\, {b_{n-1}}}^{\prime (0)}}] =0 \\
&& [ad_{\mathcal{X}_{{a_1}\,\ldots\, {a_{n-1}}}^{\prime (0)}},
ad_{\mathcal{Y}_{{b_1}\,\ldots\, {b_{n-2}} \, {v_{n-1}}}^{\prime (1)}}]=
\frac12 \epsilon_{{a_1}\,\ldots\, {a_{n-1}}\, {v_{n-1}}}{}^{v}
ad_{(X^\prime_{b_1}, \ldots, X^\prime_{b_{n-2}}, X_v)}  \nonumber\\
&& \qquad  - \frac12 \sum_{i=1}^{n-1} \epsilon_{{b_1}\,\ldots\, {b_{n-2}} \, {v_{n-1}} \, {a_i}}{}^{v}
ad_{(X_{a_1}^\prime, \ldots, {X}^\prime_{a_{i-1}}, X^\prime_v,
{X}^\prime_{a_{i+1}}, \ldots, X^\prime_{a_{n-1}} )}
 \quad  \in \mathcal{W}^{\prime(1)}       \label{LieA4cn+1,1}     \\
&& [ad_{\mathcal{X}_{{a_1}\,\ldots\, {a_{n-2}} \, {u_{n-1}}}^{\prime (1)}},
ad_{\mathcal{Y}_{{b_1}\,\ldots\, {b_{n-2}} \, {v_{n-1}}}^{\prime (1)}}] =  \nonumber\\
&& \qquad \frac12 \left( \sum_{i=1}^{n-2} \epsilon_{{a_1}\,\ldots\, {a_{n-2}}\,{u_{n-1}}\, {b_i}}{}^{v}
\underbrace{ad_{(X^\prime_{b_1}, \ldots, {X}^\prime_{b_{i-1}}, X^\prime_v,
{X}^\prime_{b_{i+1}}, \ldots, X^\prime_{b_{n-2}}, X^\prime_{v_{n-1}})}}_{=0}
- \underbrace{[(a,\, u) \leftrightarrow (b,\, v)]}_{=0} \right)= 0 \qquad \qquad \label{LieA4cn+1,1,1}
\end{eqnarray}

 The {\it r.h.s.} of eq.~\eqref{LieA4cn+1,1} may be non-zero only
if $n-2$ of the $a$ indices are equal to $n-2$ of the $b$ indices.
${\mathcal{W}^{\prime(0)}} \subset \textrm{Lie}\,\mathfrak{G}_c$ is
an abelian one-dimensional subalgebra $so(2)$ that acts on the
$2(n-1)$-dimensional abelian ideal $\mathcal{W}^{\prime(1)}\subset
\textrm{Lie}\,\mathfrak{G}_c$, which may be split as the sum of two
$(n-1)$-dimensional abelian subalgebras $\langle
ad_{(X^\prime_{a_1}, \ldots, {X}^\prime_{a_{n-2}}, X^\prime_n)}
\rangle \oplus \langle ad_{(X^\prime_{b_1}, \ldots,
{X}^\prime_{b_{n-2}}, X^\prime_{n+1})} \rangle$, where
$X^\prime_{n}$ and $X^\prime_{n+1}$ are the basis of $\mathfrak{V}$
(eq.~\eqref{Anspl}), on which $\mathcal{W}^{\prime (0)}$ acts
(eq.~\eqref{LieA4cn+1,1}) by rotating the $(X^\prime_{n},
X^\prime_{n+1})$ plane. Thus, Lie$\,(A_{n+1})_c$ has the semidirect
structure Lie$\,(A_{n+1})_c={{\mathcal{W}^{\prime (1)}} \, {\supset
\!\!\!\!\!\! \raisebox{1.5pt} {\tiny +}} \;\,{\mathcal{W}^{\prime
(0)}}} $ and is the $(2n-1)$-dimensional Lie algebra
$(Tr_{n-1}\oplus Tr_{n-1}) \,{\supset \!\!\!\!\!\! \raisebox{1.5pt}
{\tiny +}} \;\, so(2)$, where $so(2)$ rotates the two abelian
subalgebras (translations $Tr_{n-1}$) in the abelian Lie ideal
$\mathcal{W}^{\prime(1)}$.

We may also look here at the $n=2$ Lie algebra case. This gives
Lie$(A_3)_c = Tr_{2} \,{\supset \!\!\!\!\!\! \raisebox{1.5pt} {\tiny
+}} \;\, so(2)$, again $E_2$. This is not surprising: the centre of
$E_2$ is trivial and, since the inner derivations of a Lie algebra
$\mathfrak{g}$ are given by $\mathfrak{g}/Z(\mathfrak{g})$, we have
$(A_3)_c=E_2=\mathrm{InDer}\,(E_2)\equiv \mathrm{Lie}\,(A_3)_c$.

\section{On Lie$\,\mathfrak{G}_c$ and the contractions
(Lie$\,\mathfrak{G})_c$ of Lie$\,\mathfrak{G}$}
\label{sec.contLie3}

In Sec.~\ref{Sec.Gc} we have studied the general structure of $\mathfrak{G}_c$ and Lie$\,\mathfrak{G}_c$.
It is natural to ask ourselves whether there is any relation between Lie$\,\mathfrak{G}_c$ and
some contraction $(\textrm{Lie}\,\mathfrak{G})_c$ of the Lie algebra
Lie$\,\mathfrak{G}$ associated with the FA $\mathfrak{G}$, or,
equivalently, under which circumstances one may consider some kind
of relation for the Lie algebras in the lower {\it r.h.s.} of
the diagram
\begin{equation}\label{diagram}
\begin{array}{rcl}
\mathfrak{G} \quad & \longrightarrow & \qquad \qquad  \textrm{Lie}\,\mathfrak{G}  \\
{contr.\, limit}\; \left\downarrow\begin{array}{c}
    \\
    \\
   \end{array}
   \right. &  & \qquad \qquad \quad \left\downarrow\begin{array}{c}
    \\
    \\
   \end{array}
   \right.   {contr.\, limit} \\
  \mathfrak{G}_c \quad & \longrightarrow & \textrm{Lie}\, \mathfrak{G}_c \;\; ; \;\;
  (\textrm{Lie}\,\mathfrak{G})_c  \\
\end{array}
\end{equation}
for some contraction of Lie$\,\mathfrak{G}$. Note that we may not
expect the closure of the diagram, because Lie$\,\mathfrak{G}_c$ is
the algebra of derivations of $\mathfrak{G}_c$, while
$(\textrm{Lie}\,\mathfrak{G})_c$ is a contraction of an ordinary Lie
algebra determined by the inner derivations of $\mathfrak{G}$ and
not related with the adjoint derivations of $\mathfrak{G}_c$.
Further, there is a mismatch among the dimensions of
$(\mathrm{Lie}\,\mathfrak{G})_c$ and Lie$\,\mathfrak{G}_c$ since the
inner derivations associated with the $\mathcal{X}^{\prime (r)} \in
\wedge^{n-1} \mathfrak{G}_c$ for $r\ge 2$ are trivial by
eq.~\eqref{str-ad5} and then $\mathcal{W}^{\prime (r)}=0$ in
$\textrm{Lie}\,\mathfrak{G}_c\,$ for $r\ge 2$. Thus, $\mathrm{dim}\,
(\mathrm{Lie}\,\mathfrak{G}) = \mathrm{dim}\,
(\mathrm{Lie}\,\mathfrak{G})_c \neq \mathrm{dim}\,
\mathrm{Lie}\,\mathfrak{G}_c$, and the diagram~\eqref{diagram} does
not close. However, in the case of simple FAs, the comparison of
$\textrm{Lie}\, \mathfrak{G}_c$ and $(\textrm{Lie}\,\mathfrak{G})_c$
is simpler since for $\mathfrak{G}$ $ad$ is injective (see
Secs.~\ref{3lie}, \ref{c:nlie}) and the
$ad_{\mathcal{X}_{{i_1}\,\ldots\, {i_{n-1}}}}$ derivations determine
a basis of Lie$\,A_{n+1}$. We shall therefore restrict ourselves to
this case, and show how $\textrm{Lie}\, \mathfrak{G}_c$ and various
contractions $(\textrm{Lie}\,\mathfrak{G})_c$,
$\mathfrak{G}=A_{n+1}$, $n>2$, may be related.

To look into the problem we first notice that, given a FA
$\mathfrak{G}=\mathfrak{G}_0 \oplus \mathfrak{V}$
as a vector space, the splitting of Lie$\,\mathfrak{G}$ defined by the vector subspaces
\begin{equation}\label{splittingnLie}
{\mathcal{W}^{ (r)}}=\langle ad_{\mathcal{X}^{(r)}} \rangle \; , \quad  ad_{\mathcal{X}^{(r)}}=
ad_{(X_{a_1}, \ldots, X_{a_{n-r-1}}, X_{u_{n-r}}, \ldots, X_{u_{n-1}})} \; , \quad
X_{a_i} \in \mathfrak{G}_0 \, , \; X_{u_i}\in \mathfrak{V} \; ,
\end{equation}
allows us to perform a generalized contraction of $\mathrm{Lie}\,\mathfrak{G}$
in the sense of Weimar-Woods (W-W) \cite{Wei:00}.
The reason is that the splitting of Lie$\,\mathfrak{G}=\bigoplus \mathcal{W}^{ (r)}$
does not only say that ${\mathcal{W}^{ (0)}}$ is a subalgebra
of Lie$\,\mathfrak{G}$; the $\mathfrak{G}_0$
Filippov subalgebra condition $f_{a_1 \ldots  a_n}{}^u =0$ gives for
the Lie$\,\mathfrak{G}$ commutators the structure
\begin{equation}\label{W-W3}
[ad_{\mathcal{X}^{(r)}}, ad_{\mathcal{Y}^{(s)}}]\in \bigoplus
\langle ad_{\mathcal{Z}^{(t)}}\rangle \; , \quad
t\le r+s \,, \quad r,\,s,\,t\,=\, 0,\ldots, (n-1) \quad
\end{equation}
{\it i.e.},
\begin{equation}
\nonumber
\label{W-W3c}
[\mathcal{W}^{(r)}, \mathcal{W}^{(s)}]\subset \bigoplus  {\mathcal{W}^{(t)}} \; , \quad
t\le r+s \quad ,
\end{equation}
(proved in the Appendix), which is precisely the general condition needed to perform a
generalized contraction of Lie algebras in the sense of
Weimar-Woods (W-W) \cite{Wei:00}. This is defined as follows.
Let the vector space of a Lie algebra split as $\mathfrak{g}=\oplus \mathfrak{v}_p$, $p=0,1,\ldots,m$.
Let the subset of basis generators $X$ of
$\mathfrak{g}$ generating each subspace $\mathfrak{v}_p$ be redefined by $X
\rightarrow X^\prime = \epsilon^{p}X$ when $X\in \mathfrak{v}_p$. Then, a W-W Lie algebra
contraction (the limit $\epsilon \rightarrow 0$) exists iff the splitting of
$\mathfrak{g}$ above is such that $[\mathfrak{v}_p,\mathfrak{v}_q]\subset \oplus_s \mathfrak{v}_s$,
where $s$ runs over all the values for which $s \leq p + q$.
In the present Lie$\,\mathfrak{G}$ case, the contracted algebra $(\mathrm{Lie}\,\mathfrak{G})_{W-W}$ is
obtained by the reparametrization  $ad^\prime_{\mathcal{X}^{(r)}}=\epsilon^r ad_{\mathcal{X}^{(r)}}$ and
the limit $\epsilon \rightarrow 0$.

\subsection{Contractions of Lie$\,A_4$}

The contractions of $A_4=\mathfrak{G}_0 \oplus \mathfrak{V}$ with
respect to its two types of non-trivial subalgebras $\mathfrak{G}_0 \subset A_4$
and their associated Lie$\,(A_4)_c$ algebras were given
in Sec.~\ref{sec.A4c}. We consider here the contractions of the corresponding
Lie$\,A_4=\bigoplus_{r=0}^2 \mathcal{W}^{(r)}$, where as usual $r$
indicates the number of generators of the basis of
$\mathfrak{V}$ in the elements of $\mathcal{W}^{(r)}$ as in eq.~\eqref{splittingnLie}.

As a third case, we recall the IW contraction with respect to the
subalgebra $so(3)\subset \mathrm{Lie}\,A_4$, generated by the
elements in the first line in eq.~\eqref{so(3)}, and corresponding
to $\mathcal{W}^{(1)}$ in the splitting \eqref{splittingLieG3} of
its vector space, $\mathcal{W}^{(1)}\oplus \mathcal{W}^{(2)}$. Since
Lie$\, A_4$ is semisimple, there is another well known contraction,
also mentioned in Sec.~\ref{sec.third case}.

\subsubsection{First case: $\mathfrak{G}_0$ one-dimensional}

In this case $(A_4)_c$ is a four-dimensional abelian algebra and
hence Lie$\,(A_4)_c$ reduces to the trivial endomorphism (Sec.~\ref{sec.A4c}).
The IW contraction $(\mathrm{Lie}\,A_4)_c$ of $\mathrm{Lie}\,A_4=so(4)$
with respect to the trivial subalgebra
${\mathcal{W}^{(0)}}=\langle ad_{(X_4,X_4)} \rangle = \{0\}$,
associated to $\mathfrak{G}_0= \langle X_{a_4}\rangle  \subset A_4$
is obviously a six-dimensional abelian algebra; in this extreme case,
dim$\,(\mathrm{Lie}\,A_4)_c -\mathrm{dim}\,\mathrm{Lie}\,(A_4)_c =6$.

The W-W contraction for the splitting \eqref{splittingnLie} gives again a six-dimensional
abelian algebra.

\subsubsection{Second case: $(\mathrm{Lie}\,A_4)_c$, $\mathfrak{G}_0$ bidimensional}\label{IWsecondcase}

Since $n=3$ there are three types of $\mathcal{W}^{(r)}$ spaces, $r=0,1,2$.
Labelling the elements $ad_{(X_{i}, X_{j})}$ as usual, the Lie$\,A_4$ commutators are given by
\begin{equation}
\begin{array}{ll} \label{lieA4_2}
\;\;[ad_{(X_{a_1}, X_{a_2})},ad_{(X_{b_1}, X_{b_2})}]=0
& \Rightarrow [{\mathcal{W}^{(0)}},{\mathcal{W}^{(0)}}]=0 \\
\;\;[ad_{(X_{a_1}, X_{a_2})},ad_{(X_{b_1}, X_{u_1})}]=
\frac12 \epsilon_{a_1a_2u_1}{}^{u_2} ad_{(X_{b_1},X_{u_2})} & \\
\;\; \qquad -\frac12 \epsilon_{b_1 u_1 a_1}{}^{u_2} ad_{(X_{u_2},X_{a_2})}
-\frac12 \epsilon_{b_1u_1a_2}{}^{u_2} ad_{(X_{a_1},X_{u_2})}
&\Rightarrow [{\mathcal{W}^{(0)}},{\mathcal{W}^{(1)}}] \subset {\mathcal{W}^{(1)}} \\
 \left. \begin{array}{l} [ad_{(X_{a_1}, X_{u_1})},
 ad_{(X_{a_2}, X_{u_2})}]= 0\,, \; a_1 \neq a_2, u_1 \neq u_2\\
{}[ad_{(X_{a_1}, X_{u})},ad_{(X_{a_2}, X_{u})}]= \epsilon_{a_1u_1a_2}{}^{v} ad_{(X_{v},X_{u})} \\
{}[ad_{(X_{a}, X_{u_1})},ad_{(X_{a}, X_{u_2})}]= \epsilon_{a u_1u_2}{}^{b} ad_{(X_{a},X_{b})} \\
\end{array} \right\} & \Rightarrow [{\mathcal{W}^{(1)}},{\mathcal{W}^{(1)}}]
\subset {\mathcal{W}^{(0)}} \oplus {\mathcal{W}^{(2)}} \\
\;\; [ad_{(X_{a_1}, X_{a_2})},ad_{(X_{u_1}, X_{u_2})}]= 0
&\Rightarrow [{\mathcal{W}^{(0)}},{\mathcal{W}^{(2)}}]=0 \\
\;\;[ad_{(X_{u}, X_{a_1})},ad_{(X_{u}, X_{v})}]=
\epsilon_{u a_1 v}{}^{a_2} ad_{(X_{u},X_{a_2})}
&\Rightarrow [{\mathcal{W}^{(1)}},{\mathcal{W}^{(2)}}] \subset {\mathcal{W}^{(1)}} \\
\; \;[ad_{(X_{v_1}, X_{v_2})},ad_{(X_{u_1}, X_{u_2})}]= 0
& \Rightarrow  [{\mathcal{W}^{(2)}},{\mathcal{W}^{(2)}}]=0 \quad . \\
\end{array}
\end{equation}\\

{\bf - IW contraction, $(\textrm{Lie}\,A_4)_{IW}$}

We contract with respect to $\mathcal{W}^{(0)}$, the one-dimensional
subalgebra generated by $ad_{(X_{a_1}, X_{a_2})}$.
The reparametrization $ad^\prime_{\mathcal{X}^{(0)}} =
ad_{\mathcal{X}^{(0)}}, \, ad^\prime_{\mathcal{X}^{(r)}} =
\epsilon ad_{\mathcal{X}^{(r)}}, \, r=1,2$, and
the limit $\epsilon\rightarrow 0$ gives the contracted Lie algebra $(\textrm{Lie}\,A_4)_{IW}$
\begin{equation}
\begin{array}{ll}\label{LieA4IW}
[ad^\prime_{(X_{a_1}, X_{a_2})},ad^\prime_{(X_{b_1}, X_{b_2})}]=0 &
\Rightarrow [{\mathcal{W}^{(0)}},{\mathcal{W}^{(0)}}]=0 \\
{}[ad^\prime_{(X_{a_1}, X_{a_2})},ad^\prime_{(X_{b_1}, X_{u_1})}]
= \frac12 \epsilon_{a_1a_2u_1}{}^{u_2} ad^\prime_{(X_{b_1},X_{u_2})}  & \\
\;\; \qquad -\frac12 \epsilon_{b_1 u_1 a_1}{}^{u_2} ad^\prime_{(X_{u_2},X_{a_2})}-
\frac12 \epsilon_{b_1u_1a_2}{}^{u_2} ad^\prime_{(X_{a_1},X_{u_2})} &
\Rightarrow [{\mathcal{W}^{(0)}},{\mathcal{W}^{(1)}}] \subset {\mathcal{W}^{(1)}} \\
{}[ad^\prime_{(X_{a_1}, X_{u_1})},ad^\prime_{(X_{a_2}, X_{u_2})}]= 0\, &
\Rightarrow {}[{\mathcal{W}^{(1)}},{\mathcal{W}^{(1)}}] =0 \\
{}[ad^\prime_{\mathcal{X}^{(2)}}, ad^\prime_{\mathcal{Y}^{(r)}}]=0\;, \quad r=0,1,2  &
\Rightarrow  [{\mathcal{W}^{(2)}},{\mathcal{W}^{(r)}}]=0 \;, \quad r=0,1,2 \; ,\\
\end{array}
\end{equation}
where we are using the same notation $\mathcal{W}^{(r)}$ to refer now to
the subspaces of the {\it contracted} $(\textrm{Lie}\,A_4)_{IW}$
algebra. Thus, the contraction $(\textrm{Lie}\,A_4)_{IW}$ contains
Lie$\,(A_4)_c$ as a subalgebra (see
eqs.~\eqref{LieA4c-1}-\eqref{LieA4c-4}), but contains an extra
commuting generator $ad^\prime_{(X_{u_1}, X_{u_2})}$ that extends
Lie$\,(A_4)_c$ by a direct sum:
 $(\textrm{Lie}\,A_4)_{IW}=
 (Tr_{n-1}\oplus Tr_{n-1}) \,{\supset \!\!\!\!\!\! \raisebox{1.5pt}
{\tiny +}} \;\, so(2)\oplus \mathcal{W}^{(2)}=
(\textrm{Lie}\,A_4)_{c}\oplus\mathcal{W}^{(2)}$;
 dimensionally, $6=5+1$ . This result also follows from eq.~\eqref{so(4)} by
 contracting with respect to $Z_3\equiv ad_{\mathcal{X}_{12}}$ and
 with $\mathcal{W}^{(2)}$ generated by $Y_3\equiv ad_{\mathcal{X}_{43}}$.\\

{\bf - W-W generalized contraction, $(\textrm{Lie}\,A_4)_{W-W}$}

This is obtained by the reparametrizations
$ad^\prime_{\mathcal{X}^{(r)}} = \epsilon^r ad_{\mathcal{X}^{(r)}}, \, r=0,1,2$.
The $\epsilon\rightarrow 0$
limit gives $(\mathrm{Lie}\, A_4)_{W-W}$ as
\begin{equation}
\begin{array}{ll}
\;\;[ad^\prime_{(X_{a_1}, X_{a_2})},ad^\prime_{(X_{b_1}, X_{b_2})}]=0 &
\Rightarrow [{\mathcal{W}^{(0)}},{\mathcal{W}^{(0)}}]=0 \\
\;\;[ad^\prime_{(X_{a_1}, X_{a_2})},ad^\prime_{(X_{b_1}, X_{u_1})}]
= \frac12 \epsilon_{a_1a_2u_1}{}^{u_2} ad^\prime_{(X_{b_1},X_{u_2})} & \\
\;\; \qquad -\frac12 \epsilon_{b_1 u_1 a_1}{}^{u_2} ad^\prime_{(X_{u_2},X_{a_2})}-
\frac12 \epsilon_{b_1u_1a_2}{}^{u_2} ad^\prime_{(X_{a_1},X_{u_2})} &
\Rightarrow [{\mathcal{W}^{(0)}},{\mathcal{W}^{(1)}}] \subset {\mathcal{W}^{(1)}} \\
 \left. \begin{array}{l} [ad^\prime_{(X_{a_1}, X_{u_1})},
 ad^\prime_{(X_{a_2}, X_{u_2})}]= 0\,, \; a_1 \neq a_2, u_1 \neq u_2\\
{}[ad^\prime_{(X_{a_1}, X_{u})},ad^\prime_{(X_{a_2}, X_{u})}]
= \epsilon_{a_1ua_2}{}^{v} ad^\prime_{(X_{v},X_{u})} \\
{}[ad^\prime_{(X_{a}, X_{u_1})},ad^\prime_{(X_{a}, X_{u_2})}]= 0 \\
\end{array} \right\} & \Rightarrow [{\mathcal{W}^{(1)}},{\mathcal{W}^{(1)}}] \subset {\mathcal{W}^{(2)}} \\
\;\; [ad^\prime_{\mathcal{X}^{(r)}}, ad^\prime_{\mathcal{Y}^{ (2)}}]=0\;, \quad r=0,1,2  &
\Rightarrow  [{\mathcal{W}^{(r)}},{\mathcal{W}^{(2)}}]=0 \;, \quad r=0,1,2 \;. \\
\end{array}
\end{equation}
This is a central extension of Lie$\,(A_4)_c$ (eqs.~\eqref{LieA4c-1}-\eqref{LieA4c-4})
by the one-dimensional subalgebra $\mathcal{W}^{(2)}= \langle ad^\prime_{(X_{u_1}, X_{u_2})} \rangle$.
Thus, Lie$\,(A_4)_c$= $(\mathrm{Lie}\, A_4)_{W-W}/\mathcal{W}^{(2)}$,
and it is not a subalgebra of $(\mathrm{Lie}\,A_4)_{W-W}$.

\subsubsection{Third case}\label{sec.third case}

Consider Lie$\, A_4$ as given by the sum ${\mathcal{W}^{(1)}} \oplus
{\mathcal{W}^{(2)}}$ where ${\mathcal{W}^{(1)}}=\langle Y_1, Y_2,
Y_3 \rangle $ is a $so(3)$ subalgebra and
${\mathcal{W}^{(2)}}=\langle Z_1, Z_2, Z_3 \rangle $
(eq.~\eqref{so(3)}). The IW contraction with respect to the
${\mathcal{W}^{(1)}}$ subalgebra is the well known $6$-dimensional
Euclidean algebra $E_3$,  $(\textrm{Lie}\,A_4)_c= {\mathcal{W}^{(2)}}
\, {\supset \!\!\!\!\!\! \raisebox{1.5pt} {\tiny
+}}\;\,{\mathcal{W}^{(1)}}$ $\equiv Tr_3 \, {\supset \!\!\!\!\!\!
\raisebox{1.5pt} {\tiny +}}\;\, so(3)$.

Since Lie$\,A_4=so(3)\oplus so(3)$ is not simple,
there is of course the possibility of contracting with
respect to any of the $so(3)$ subalgebras in eq.~\eqref{so(4)tilde},
leading to $Tr_3 \oplus so(3)$.

\subsection{Contractions of Lie$\,A_{n+1}$}

In Sec.~\ref{Sec.An+1c} we have seen that the only splitting of $\mathfrak{G}$ that leads
to a non-trivial contracted Filippov algebra $(A_{n+1})_c$
requires taking $\mathfrak{G}_0$ as an abelian
subalgebra generated by $n-1$ $A_{n+1}$ basis elements
so that $\mathfrak{V}$ is generated by the remaining two,
$\mathfrak{G}_0= \langle X_a, \; a=1,\ldots,n-1 \rangle,
\quad \mathfrak{V}=\langle X_u, \; u=n,n+1 \rangle $.

Labelling as in eq.~\eqref{splittingnLie}, the commutators of
Lie$\,A_{n+1}$ for the different subspaces adopt the form:
\begin{eqnarray}
&& [ad^{(0)}_{(X_{a_1}, \ldots, X_{a_{n-1}})}, ad^{(0)}_{(X_{b_1}, \ldots, X_{b_{n-1}})}] =
\frac12 \sum_{i=1}^{n-1} \underbrace{\epsilon_{{a_1}\,\ldots\, {a_{n-1}}\, {b_i}}{}^{b}}_{=0}
ad_{(X_{b_1}, \ldots, {X}_{b_{i-1}}, X_b, {X}_{b_{i+1}}, \ldots, X_{b_{n-1}})}
\qquad \nonumber \label{liean+1c_1}\\
&& \quad -\frac12 \sum_{i=1}^{n-1} \underbrace{\epsilon_{{b_1}\,\ldots\, {b_{n-1}}\, {a_i}}{}^{b}}_{=0}
ad_{(X_{a_1}, \ldots, {X}_{a_{i-1}}, X_b, {X}_{a_{i+1}}, \ldots, X_{a_{n-1}})}=0
\\[0.3cm]
&& [ad^{(0)}_{(X_{a_1}, \ldots, X_{a_{n-1}})}, ad^{(1)}_{(X_{b_1}, \ldots, X_{b_{n-2}}, X_{v_{n-1}})}] =
\qquad \nonumber \\
&& \quad \frac12 \sum_{i=1}^{n-2} \underbrace{\epsilon_{{a_1}\,\ldots\, {a_{n-1}}\, {b_i}}{}^{b}}_{=0}
ad_{(X_{b_1}, \ldots, {X}_{b_{i-1}}, X_b, {X}_{b_{i+1}}, \ldots, X_{b_{n-2}}, X_{v_{n-1}})} \nonumber \\
&& \quad + \frac12 \underbrace{\epsilon_{{a_1}\,\ldots\, {a_{n-1}}\, {v_{n-1}}}{}^{b}}_{=0}
ad_{(X_{b_1}, \ldots, X_{b_{n-2}}, X_b)} +
\frac12 \epsilon_{{a_1}\,\ldots\, {a_{n-1}}\, {v_{n-1}}}{}^{v}
ad_{(X_{b_1}, \ldots, X_{b_{n-2}}, X_v)} \nonumber\\
&&\quad - \frac12 \sum_{i=1}^{n-1} f_{{b_1}\,\ldots\, {b_{n-2}} \, {v_{n-1}} \, {a_i}}{}^{v}
ad_{(X_{a_1}, \ldots, {X}_{a_{i-1}}, X_v, {X}_{a_{i+1}}, \ldots, X_{a_{n-1}} )} \quad
\in \mathcal{W}^{(1)} \\[0.3cm]
&& [ad^{(0)}_{(X_{a_1}, \ldots, X_{a_{n-1}})} ,
ad^{(2)}_{(X_{b_1}, \ldots, X_{b_{n-3}}, X_{v_{n-2}}, X_{v_{n-1}})}] = \qquad \nonumber \\
&& \quad \frac12 \sum_{i=1}^{n-3} \underbrace{\epsilon_{{a_1}\,\ldots\, {a_{n-1}}\, {b_i}}{}^{b}}_{=0}
ad_{(X_{b_1}, \ldots,{X}_{b_{i-1}}, X_b, {X}_{b_{i+1}},
\ldots, X_{b_{n-3}}, X_{v_{n-2}}, X_{v_{n-1}})} \nonumber \\
&& \quad + \frac12 \underbrace{\epsilon_{{a_1}\,\ldots\, {a_{n-1}}\, {v_{n-2}}}{}^{b}}_{=0}
ad_{(X_{b_1}, \ldots, X_{b_{n-3}}, X_{v_{n-1}}, X_b)}
+ \frac12 \underbrace{\epsilon_{{a_1}\,\ldots\, {a_{n-1}}\, {v_{n-2}}}{}^{v}
ad_{(X_{b_1}, \ldots, X_{b_{n-3}}, X_{v_{n-1}}, X_v)}}_{=0} \quad \nonumber \\
&& \quad + \frac12 \underbrace{\epsilon_{{a_1}\,\ldots\, {a_{n-1}}\, {v_{n-1}}}{}^{b}}_{=0}
ad_{(X_{b_1}, \ldots, X_{b_{n-3}}, X_b, X_{v_{n-2}})}
+ \frac12 \underbrace{\epsilon_{{a_1}\,\ldots\, {a_{n-1}}\, {v_{n-1}}}{}^{v}
ad_{(X_{b_1}, \ldots, X_{b_{n-3}}, X_v, X_{v_{n-2}})}}_{=0} \nonumber\\
&&\quad - \frac12 \sum_{i=1}^{n-1} f_{{b_1}\,\ldots\, {b_{n-3}} \, {v_{n-2}}\, {v_{n-1}} \, {a_i}}{}^{b}
\underbrace{ad_{(X_{a_1}, \ldots, {X}_{a_{i-1}}, X_b, {X}_{a_{i+1}},, \ldots, X_{a_{n-1}} )}}_{=0}
\nonumber\\
&&\quad - \frac12 \sum_{i=1}^{n-1} \underbrace{f_{{b_1}\,\ldots\, {b_{n-3}} \, {v_{n-2}}\,
{v_{n-1}} \, {a_i}}{}^{v}}_{=0}
ad_{(X_{a_1}, \ldots, {X}_{a_{i-1}}, X_v, {X}_{a_{i+1}}, \ldots, X_{a_{n-1}} )} \quad =0
\\[0.3cm]
&& [ad^{(1)}_{(X_{a_1}, \ldots, X_{a_{n-2}},X_{u_{n-1}})},
ad^{(1)}_{(X_{b_1}, \ldots, X_{b_{n-2}}, X_{v_{n-1}})}]  = \qquad \nonumber \\
&& \quad \frac12 \left( \sum_{i=1}^{n-2} \underbrace{\epsilon_{{a_1}\,\ldots\,
{a_{n-2}}\,{u_{n-1}}\, {b_i}}{}^{b}}_{=0} ad_{(X_{b_1}, \ldots, {X}_{b_{i-1}}, X_b, {X}_{b_{i+1}},
\ldots, X_{b_{n-2}}, X_{v_{n-1}})} \nonumber \right. \\
&& \quad + \sum_{i=1}^{n-2} \epsilon_{{a_1}\,\ldots\, {a_{n-2}}\,{u_{n-1}}\, {b_i}}{}^{v}
ad_{(X_{b_1}, \ldots, {X}_{b_{i-1}}, X_v, {X}_{b_{i+1}}, \ldots,
X_{b_{n-2}}, X_{v_{n-1}})}\nonumber \\
&& \quad \left. + \epsilon_{{a_1}\,\ldots\, {a_{n-2}}\,{u_{n-1}}\, {v_{n-1}}}{}^{b}
ad_{(X_{b_1}, \ldots, X_{b_{n-2}}, X_b)}
+  \underbrace{\epsilon_{{a_1}\,\ldots\, {a_{n-3}}\,{u_{n-1}}\, {v_{n-1}}}{}^{v}}_{=0}
ad_{(X_{b_1}, \ldots, X_{b_{n-2}}, X_v)} \right) \quad \nonumber \\
&& \quad - \left[ (a,\, u) \leftrightarrow (b,\, v) \right]\quad \in \mathcal{W}^{(0)} \oplus \mathcal{W}^{(2)}
\\[0.3cm]
&& [ad^{(1)}_{(X_{a_1}, \ldots, X_{a_{n-2}},X_{u_{n-1}})},
ad^{(2)}_{(X_{b_1}, \ldots, X_{b_{n-3}}, X_{v_{n-2}}, X_{v_{n-1}})}]= \nonumber \\
&&  \quad \frac12 \sum_{i=1}^{n-2} \underbrace{\epsilon_{{a_1}\,\ldots\, {a_{n-2}}\,{u_{n-1}}\, {b_i}}{}^{b}}_{=0}
ad_{(X_{b_1}, \ldots, {X}_{b_{i-1}}, X_b, {X}_{b_{i+1}}, \ldots, X_{b_{n-3}}, X_{v_{n-2}}, X_{v_{n-1}})}\nonumber \\
&& \quad + \frac12 \sum_{i=1}^{n-2} \epsilon_{{a_1}\,\ldots\, {a_{n-2}}\,{u_{n-1}}\, {b_i}}{}^{v}
\underbrace{ad_{(X_{b_1}, \ldots, {X}_{b_{i-1}}, X_v, {X}_{b_{i+1}}, \ldots,
X_{b_{n-3}}, X_{v_{n-2}}, X_{v_{n-1}})}}_{=0}\\
&& \quad + \frac12 \epsilon_{{a_1}\,\ldots\, {a_{n-2}}\,{u_{n-1}}\, {v_{n-2}}}{}^{b}
ad_{(X_{b_1}, \ldots, X_{b_{n-3}}, X_{v_{n-1}}, X_b)}
+ \frac12 \underbrace{\epsilon_{{a_1}\,\ldots\, {a_{n-3}}\,{u_{n-1}}\, {v_{n-2}}}{}^{v}}_{=0}
ad_{(X_{b_1}, \ldots, X_{b_{n-3}}, X_{v_{n-1}}, X_v)} \quad \nonumber \\
&& \quad + \frac12 \epsilon_{{a_1}\,\ldots\, {a_{n-2}}\,{u_{n-1}}\, {v_{n-1}}}{}^{b}
ad_{(X_{b_1}, \ldots, X_{b_{n-3}}, X_b, X_{v_{n-2}})}
+ \frac12 \underbrace{\epsilon_{{a_1}\,\ldots\, {a_{n-3}}\,{u_{n-1}}\, {v_{n-1}}}{}^{v}}_{=0}
ad_{(X_{b_1}, \ldots, X_{b_{n-3}}, X_v, X_{v_{n-2}})} \quad \nonumber\\[-0.3cm]
&& \quad - \frac12 \sum_{i=1}^{n-2} \epsilon_{{b_1}\,\ldots\, {b_{n-3}}\,v_{n-2}\,  v_{n-1}\, {a_i}}{}^{b}
ad_{(X_{a_1}, \ldots, {X}_{a_{i-1}}, X_b, {X}_{a_{i+1}}, \ldots, X_{a_{n-2}}, X_{u_{n-1}})}\nonumber \\
&& \quad - \frac12 \sum_{i=1}^{n-2}
\underbrace{\epsilon_{{b_1}\,\ldots\, {b_{n-3}}\,v_{n-2}\, v_{n-1}\, {a_i}}{}^{v}}_{=0}
ad_{(X_{a_1}, \ldots, {X}_{a_{i-1}}, X_v, {X}_{a_{i+1}}, \ldots, X_{a_{n-2}}, X_{u_{n-1}})}\nonumber \\
&&  \quad - \frac12 \underbrace{\epsilon_{{b_1}\,\ldots\, {b_{n-3}}\, {v_{n-2}}\,  v_{n-1} \,{u_{n-1}}}{}^{l}}_{=0}
ad_{(X_{a_1}, \ldots, X_{a_{n-2}}, X_l)}\quad\nonumber \\
&& \quad \in \mathcal{W}^{(1)}\\[0.3cm]
&& [ad^{(0)}_{(X_{a_1}, \ldots, X_{a_{n-1}})} ,
\underbrace{ad^{(3)}_{(X_{b_1}, \ldots, X_{b_{n-4}},  X_{v_{n-3}},  X_{v_{n-2}}, X_{v_{n-1}})}}_{=0}] = 0
\\[0.1cm]
&& [ad^{(r)}_{(X_{a_1}, \ldots, X_{a_{n-r-1}},X_{u_{n-r}},\ldots, X_{u_{n-1}})},
ad^{(s)}_{(X_{b_1}, \ldots, X_{b_{n-s-1}}, X_{v_{n-s}}, \ldots X_{v_{n-1}})}]= \nonumber \\
&& \quad = \frac12 \sum_{i=1}^{n-s-1} \epsilon_{{a_1}\,\ldots\, {a_{n-r-1}}\,u_{n-r}\, \ldots \, u_{n-1}\, {b_i}}{}^{l}
ad_{(X_{b_1}, \ldots, {X}_{b_{i-1}}, X_l, {X}_{b_{i+1}}, \ldots, X_{b_{n-s-1}}, X_{v_{n-s}}, \ldots, X_{v_{n-1}})}\nonumber \\
&&  \quad + \frac12 \sum_{i=n-s}^{n-1} \epsilon_{{a_1}\,\ldots\, {a_{n-r-1}}\, {u_{n-r}}\, \ldots \, u_{n-1} \,{v_i}}{}^{l}
ad_{(X_{b_1}, \ldots, X_{b_{n-s-1}}, X_{v_{n-s}}, \ldots, {X}_{v_{i-1}}, X_l, {X}_{v_{i+1}}, \ldots , X_{v_{n-1}})}\nonumber\\
&& \quad - \left[ (a,\, u,\, r) \leftrightarrow (b,\, v,\, s) \right] =0, \qquad  r+s > 3\, . \quad \label{liean+1c_r}
\end{eqnarray}
where the constants $\epsilon_{{l_1}\,\ldots\, {l_{n}}}{}^{j}$ are zero if they contain more than $n-1$ indices $l_i\in I_0$ (cf.~\eqref{splittingnLieI}) or more than 2 indices $l_i\in I_1$; the inner endomorphisms
$ad_{(X_{l_1}, \ldots, X_{l_{n-1}})}$ are zero if they contain more than two indices
$l_i\in I_1$. For $n=3$, the above expressions reduce to eqs.~\eqref{lieA4_2}.

Since $\mathrm{dim}\,\mathcal{W}^{(0)}=\left(
  \begin{array}{c}
    n-1 \\
    n-1 \\
  \end{array}
\right) =1 \, , \;
\mathrm{dim}\,\mathcal{W}^{(1)}=2\left(
  \begin{array}{c}
    n-1 \\
    n-2 \\
  \end{array}
\right) =2(n-1) \, , \;
\mathrm{dim}\,\mathcal{W}^{(2)}=\left(
  \begin{array}{c}
    n-1 \\
    n-3 \\
  \end{array}
\right) =\frac12 (n-1)(n-2) \, , \;
\mathrm{dim}\,\mathcal{W}^{(r)}=0, \; r>2$, we check that $\mathrm{dim}\,\mathcal{W}^{(0)}+\mathrm{dim}\,\mathcal{W}^{(1)}+\mathrm{dim}\,\mathcal{W}^{(2)}=\left(
  \begin{array}{c}
    n+1 \\
    2 \\
  \end{array}
\right) = \mathrm{dim}\,\mathrm{Lie}\,A_{n+1}\, .$

\subsubsection{IW contraction $(\mathrm{Lie}\,A_{n+1})_{IW}$ of Lie$\,A_{n+1}$}
The contraction $(\mathrm{Lie}\,A_{n+1})_{IW}$, obtained by the reparametrization
$ad^\prime_{\mathcal{X}^{(0)}} = ad_{\mathcal{X}^{(0)}}\,$
$( \langle ad_{\mathcal{X}^{(0)}} \rangle  =
\mathrm{Lie}\, \mathfrak{G}_0 ), \, ad^\prime_{\mathcal{X}^{(r)}} = \epsilon
ad_{\mathcal{X}^{(r)}}, \, r=1,\ldots,n-1$, is given by
\begin{eqnarray}
&& [ad^{\prime(0)}_{(X_{a_1}, \ldots, X_{a_{n-1}})}, ad^{\prime(0)}_{(X_{b_1},
\ldots, X_{b_{n-1}})}] =0
\label{LieAnIW1}
\\
&& [ad^{\prime(0)}_{(X_{a_1}, \ldots, X_{a_{n-1}})}, ad^{\prime(1)}_{(X_{b_1}, \ldots, X_{b_{n-2}}, X_{v_{n-1}})}] =
\frac12 \epsilon_{{a_1}\,\ldots\, {a_{n-1}}\, {v_{n-1}}}{}^{v}
ad^\prime_{(X_{b_1}, \ldots, X_{b_{n-2}}, X_v)} \nonumber\\
&&\quad - \frac12 \sum_{i=1}^{n-1} f_{{b_1}\,\ldots\, {b_{n-2}} \, {v_{n-1}} \, {a_i}}{}^{v}
ad^\prime_{(X_{a_1}, \ldots, {X}_{a_{i-1}}, X_v, {X}_{a_{i+1}}, \ldots, X_{a_{n-1}})} \quad
\in \mathcal{W}^{(1)} \qquad
\label{LieAnIW2} \\
&& [ad^{\prime(0)}_{(X_{a_1}, \ldots, X_{a_{n-1}})} ,
ad^{\prime(2)}_{(X_{b_1}, \ldots, X_{b_{n-3}}, X_{v_{n-2}}, X_{v_{n-1}})}] =0
\label{LieAnIW3} \\
&& [ad^{\prime(1)}_{(X_{a_1}, \ldots, X_{a_{n-2}},X_{u_{n-1}})},
ad^{\prime(1)}_{(X_{b_1}, \ldots, X_{b_{n-2}}, X_{v_{n-1}})}]  =  0
\label{LieAnIW4} \\
&& [ad^{\prime(r)}_{(X_{a_1}, \ldots, X_{a_{n-r-1}},X_{u_{n-r}},\ldots, X_{u_{n-1}})},
ad^{\prime(s)}_{(X_{b_1}, \ldots, X_{b_{n-s-1}}, X_{v_{n-s}}, \ldots X_{v_{n-1}})}] =0 \; ,
\; r+s > 2
\label{LieAnIW5}
\end{eqnarray}
(in fact, $r+s\geq 2$, see eqs.~\eqref{LieAnIW3}, \eqref{LieAnIW4}), and
generalizes the $(\mathrm{Lie}\,A_4)_{IW}$ case of Sec.~\ref{IWsecondcase}.
We see that eqs.~\eqref{LieAnIW1}-\eqref{LieAnIW5} give the
Lie$\,(A_{n+1})_c$ algebra plus the $\left(
  \begin{array}{c}
    n-1 \\
    n-3 \\
  \end{array}
\right)$ abelian algebra $\mathcal{W}^{(2)}$, that is,
$(\mathrm{Lie}\,A_{n+1})_{IW}= (Tr_{(n-1)}\oplus Tr_{(n-1)}) \,
{\supset \!\!\!\!\!\! \raisebox{1.5pt} {\tiny +}} \;\, so(2) \oplus \mathcal{W}^{(2)}$.
Further, dim$\,(\mathrm{Lie}\,A_{n+1})_{IW} =
\mathrm{dim}\,\mathcal{W}^{(1)} + \mathrm{dim}\,\mathcal{W}^{(0)}+\mathrm{dim}\,\mathcal{W}^{(2)} =
\mathrm{dim}\,\left[(\mathrm{Lie}\,A_{n+1})=so(n+1)\right]$.
For $n=3$ the above commutators lead to eqs.~\eqref{LieA4IW}.

\subsubsection{W-W contraction $(\mathrm{Lie}\,A_{n+1})_{W-W}$ of Lie$\,A_{n+1}$}

The W-W reparametrization is now
$ad^\prime_{\mathcal{X}^{(r)}} = \epsilon^r ad_{\mathcal{X}^{(r)}}$
and in the limit $\epsilon\rightarrow 0$, eqs.~\eqref{liean+1c_1}-\eqref{liean+1c_r}
lead to the same $n$-brackets as in eqs.~\eqref{LieAnIW1}-\eqref{LieAnIW5},
but for \eqref{LieAnIW4} which is replaced by
\begin{eqnarray}
&& [ad^{\prime(1)}_{(X_{a_1}, \ldots, X_{a_{n-2}},X_{u_{n-1}})},
ad^{\prime(1)}_{(X_{b_1}, \ldots, X_{b_{n-2}}, X_{v_{n-1}})}]  = \nonumber\\
&& \quad \frac12 \epsilon_{{a_1}\,\ldots\, {a_{n-2}}\,{u_{n-1}}\, {b_i}}{}^{v}
ad^\prime_{(X_{b_1}, \ldots, {X}_{b_{i-1}}, X_v, {X}_{b_{i+1}}, \ldots, X_{b_{n-2}}, X_{v_{n-1}})}
\nonumber\\
&& \quad - \frac12 \epsilon_{{b_1}\,\ldots\, {b_{n-2}}\,{v_{n-1}}\, {a_i}}{}^{v}
ad^\prime_{(X_{a_1}, \ldots, {X}_{a_{i-1}}, X_v, {X}_{a_{i+1}}, \ldots, X_{a_{n-2}}, X_{u_{n-1}})}
\in \mathcal{W}^{(2)}
\end{eqnarray}
which indicates that $(\mathrm{Lie}\,A_{n+1})_{W-W}$ is a central extension of the $(2n-1)$-dimensional
Lie$\,(A_{n+1})_c$ (see below eq.~\eqref{spLieAn+1c}) by the $\left(
  \begin{array}{c}
    n-1 \\
    n-3 \\
  \end{array}
\right)$-dimensional abelian algebra $\mathcal{W}^{(2)}=\langle
ad^\prime_{\mathcal{X}^{(2)}} \rangle $ so that
$(\mathrm{Lie}\,A_{n+1})_{W-W} /
\mathcal{W}^{(2)}=\mathrm{Lie}\,(A_{n+1})_{c}$ as given by
eqs.~\eqref{LieA4cn+1,0,0}-\eqref{LieA4cn+1,1,1} (Lie$\,(A_{n+1})_c$
is not a subalgebra of $(\mathrm{Lie}\,A_{n+1})_{W-W}$). Of course,
$\left(
  \begin{array}{c}
    n+1 \\
    2 \\
  \end{array}
\right) -
\left(
  \begin{array}{c}
    n-1 \\
    n-3 \\
  \end{array}
\right)
=\mathrm{dim}\, \mathrm{Lie}\, (A_{n+1})_c$.

\section{Conclusions}
\label{conclusions}

We have introduced in this paper the contractions of Filippov
algebras and given the non-trivial IW-type
contractions of the $A_{n+1}$ simple FAs to illustrate the
procedure. As it is for the Lie algebras case, the contraction of a
FA $\mathfrak{G}$ has to be done with respect to a subalgebra
$\mathfrak{G}_0$ and has the semidirect FA structure
$\mathfrak{G}_c={\mathfrak{V}}\, {\supset \!\!\!\!\!\!
\raisebox{1.5pt} {\tiny +}} \;\,{\mathfrak{G}_0}$, where
$\mathfrak{V}$ is a FA abelian ideal of $\mathfrak{G}_c$.

We have also considered the Lie algebra Lie$\,\mathfrak{G}_c$
associated with a FA contraction $\mathfrak{G}_c$, and the contractions
$(\mathrm{Lie}\,\mathfrak{G})_c$ of the Lie algebra associated with
the uncontracted FA $\mathfrak{G}$, and compared them in the simple
$\mathfrak{G}=A_{n+1}$ case. We have seen that the IW or W-W
contractions $(\mathrm{Lie}\,A_{n+1})_{IW}$,
$(\mathrm{Lie}\,A_{n+1})_{W-W}$
are either a trivial or a central extension of the
Lie algebra $\mathrm{Lie}\,(A_{n+1})_c$ associated with the
non-trivial contraction of the simple Filippov algebra $A_{n+1}$.

All the examples in this paper have dealt with simple FAs. It is
clear that, for semisimple FAs, a contraction that only affects the
generators of a single ideal will not modify the others since they
remain as spectators of the contraction process. But, already for
$n=2$, it is possible to define IW contractions of Lie algebras
which have direct sum structure by using a basis that
contains generators involving a combination of those of
different algebras in the direct sum, be these simple ones or
not. The result of a contraction of this type is a Lie algebra that
does no longer have the original direct sum structure of the
uncontracted one (these contractions are sometimes
called `unconventional', `exotic' or even `generalized', although
they are ordinary, standard IW contractions).
This explains why the well known eleven dimensional,
centrally extended Galilei group may be obtained by a contraction
of the direct product of the Poincar\'e group and a $U(1)$ factor
(see \cite{Sal:61} and \cite{Al-Az:85, CUP} for the `generating
cohomology' properties of these contractions). Other physical examples
of contractions of this type have been considered in \cite{C-J:92},
in \cite{AIPV:03} in the context of expansions of Lie algebras (a process
\cite{H-S:03,AIPV:03,AIPV:04-07} that is not dimension preserving in
general but that includes IW contractions as a particular case) and,
very recently, in \cite{Hor-Mar-Sti:10, Luk:10}.

In our $n>2$ case, this type of contractions may have a bearing
for FAs. It is well known that it is not easy to find explicit examples
of FAs beyond the semisimple ones, one of the reasons being the lack of
associativity: the Filippov bracket is not constructed from
associative products of its $n$ entries ({\it cf.} Refs.~\cite{CMP,HW95};
see ref.~\cite{review} for further discussion). The above type
of contractions, applied to direct sums of FAs, would lead to other
non-trivial examples of FAs. Note that here, however, we would
be dealing -as throughout this paper- with Filippov {\it algebras} only;
for $n>2$, there is no  `Filippov group' manifold
structure and no vector fields associated with
FA  generators that could act on it.

    Finally, one might think of applying the above contraction
scheme to some physical situa\-tion. As an exercise, we have tried
it on the original Bagger-Lambert-Gustavsson
$A_4$ model \cite{Ba-La:06-07,Gus:07,Gustav:08} of coincident M2 branes
(see {\it e.g.} \cite{review} for further references),
but the resulting Lagrangian becomes trivial: the Chern-Simons term
disappears and, further, it reduces to the free kinetic terms.

\subsection*{Acknowledgments}

This work has been partially supported by research grants from the
Spanish Ministry of Science and Innovation (FIS2008-01980,
FIS2005-03989), the Junta de Castilla y Le\'on (VA013C05) and EU
FEDER funds. M.P. would like to thank the Spanish Ministerio de
Ciencia e Innovaci\'on and the Fulbright Program for the
MICINN-Fulbright postdoctoral research fellowship he held whilst this
work was being carried out, and the Department of Physics and
Astronomy at Stony Brook and specially Martin
Ro\v{c}ek for their kind hospitality.

\appendix

\section{On the graded W-W structure of the splitting
Lie$\,\mathfrak{G}=\mathcal{W}^{(0)}\oplus \cdots \oplus \mathcal{W}^{(n-1)}$}

Let $\mathfrak{G}=\mathfrak{G}_0 \oplus \mathfrak{V}$ as a vector space, and let $\mathcal{W}^{(r)}=
\langle ad_{\mathcal{X}_{a_1 \ldots a_{n-r-1} u_{n-r} \ldots u_{n-1}}}\rangle $ the
Lie$\,\mathfrak{G}$ subspaces generated by the elements
$ad_{\mathcal{X}_{a_1 \ldots a_{n-r-1} u_{n-r} \ldots u_{n-1}}}$, where the
superindex $r$ indicates the number of generators
$X_{u_{n-r}}, \ldots, X_{u_{n-1}}$ of the basis of $\mathfrak{V}$ in the fundamental object
$\mathcal{X}$ in $ad_{\mathcal{X}}$.

In terms of the structure constants of the FA $\mathfrak{G}$ and using this splitting, the
Lie$\,\mathfrak{G}$ algebra commutators are
\begin{eqnarray}
&& [ad_{\mathcal{X}_{{a_1}\,\ldots\, {a_{n-r-1}}\,u_{n-r}\, \ldots \, u_{n-1}}^{(r)}},
ad_{\mathcal{Y}_{{b_1}\,\ldots\, {b_{n-s-1}}\,v_{n-s}\, \ldots \, v_{n-1}}^{(s)}}] =  \nonumber\\
&& \quad = \frac12 ad_{\left[(X_{a_1}, \ldots, X_{a_{n-r-1}},X_{u_{n-r}},\ldots, X_{u_{n-1}})
\cdot (X_{b_1}, \ldots, X_{b_{n-s-1}}, X_{v_{n-s}}, \ldots X_{v_{n-1}}) \right.} \nonumber \\
&&\qquad  {}_{\left. -  (X_{b_1}, \ldots, X_{b_{n-s-1}}, X_{v_{n-s}}, \ldots X_{v_{n-1}})\cdot
(X_{a_1}, \ldots, X_{a_{n-r-1}},X_{u_{n-r}},\ldots, X_{u_{n-1}})\right]} = \nonumber\\
&& \quad = \frac12 \sum_{i=1}^{n-s-1} f_{{a_1}\,\ldots\, {a_{n-r-1}}\,u_{n-r}\, \ldots \, u_{n-1}\, {b_i}}{}^{l}
ad_{(X_{b_1}, \ldots, {X}_{b_{i-1}}, X_l, {X}_{b_{i+1}},
\ldots, X_{b_{n-s-1}}, X_{v_{n-s}}, \ldots, X_{v_{n-1}})}\nonumber\\
&& \quad + \frac12 \sum_{i=n-s}^{n-1} f_{{a_1}\,\ldots\, {a_{n-r-1}}\, {u_{n-r}}\, \ldots \, u_{n-1} \,{v_i}}{}^{l}
ad_{(X_{b_1}, \ldots, X_{b_{n-s-1}}, X_{v_{n-s}},
\ldots , {X}_{v_{i-1}}, X_l, {X}_{v_{i+1}}, \ldots,   X_{v_{n-1}})}  \nonumber\\
&& \quad - \frac12 \sum_{i=1}^{n-r-1} f_{{b_1}\,\ldots\, {b_{n-s-1}}\,v_{n-s}\, \ldots \, v_{n-1}\, {a_i}}{}^{l}
ad_{(X_{a_1}, \ldots, {X}_{a_{i-1}}, X_l, {X}_{a_{i+1}},
\ldots, X_{a_{n-r-1}}, X_{u_{n-r}}, \ldots, X_{u_{n-1}})}\nonumber\\
&& \quad - \frac12 \sum_{i=n-r}^{n-1} f_{{b_1}\,\ldots\,
{b_{n-s-1}}\, {v_{n-s}}\, \ldots \, v_{n-1} \,{u_i}}{}^{l}
ad_{(X_{a_1}, \ldots, X_{a_{n-r-1}}, X_{u_{n-r}}, \ldots,
{X}_{u_{i-1}}, X_l, {X}_{u_{i+1}},  \ldots , X_{u_{n-1}})} \,.
\end{eqnarray}

\noindent (as mentioned, if $\mathfrak{G}$ is not simple, not all
$ad_{(X_{a_1}, \ldots, X_{a_{n-r-1}}, X_{u_{n-r}}, \ldots
X_{u_{n-1}})}$ are independent in general). The fulfillment of the
W-W condition \eqref{W-W3} is a consequence of the dot composition
of the fundamental objects (eq.~\eqref{comp}). Indeed, the
$\mathcal{X}$'s in $ad$'s in the {\it r.h.s.} contain a maximum of
$r+s$ elements of the basis of $\mathfrak{V}$, except when $r=0$ (no
$u$ indices) or $s=0$ (no $v$ indices), where the first and third
summatories give a term with $ad_{\mathcal{Z}^{(t)}}$, $t = r+1$ or
$t=s+1$. However, the terms with $t=r+s+1$ with $r$ or $s$ equal to
zero are zero when $\mathfrak{G}_0$ is a subalgebra, $f_{a_1 \ldots
a_{n-1}}{}^u=f_{b_1 \ldots b_{n-1}}{}^u=0$, and then it follows
that $[ad_{\mathcal{X}^{(r)}}, ad_{\mathcal{Y}^{(s)}}]\in \bigoplus
\langle ad_{\mathcal{Z}^{(t)}}\rangle \; , \; t\le r+s $. Therefore,
when the above splitting of Lie$\,\mathfrak{G}$ is considered, the
W-W condition
\begin{equation}\label{WWcond}
[\mathcal{W}^{(r)}, \mathcal{W}^{(s)}]\subset \bigoplus  \mathcal{W}^{(t)} \; , \quad t\le r+s \;
\end{equation}
$(\mathrm{here},\; r,s,t=1,\ldots,n-1)$ is automatically fulfilled
when $\mathfrak{G}_0$ is subalgebra (a condition also reflected for
$r=0=s$).

\end{document}